\documentclass[11pt,a4paper]{article}
\usepackage{jheppub}

\usepackage{stmaryrd}

\def\Uq{\mathcal{U}_{q}({\mathfrak{g}})}
\def\Ua{\mathcal{U}_{q}(\widehat{\mathfrak{g}})}

\def\tth{\tilde\theta}

\def\K{{K}}
\def\R{{\mathbb R}}
\def\L{{\mathbb L}}
\def\Q{{\mathbb Q}}
\def\G{{\mathbb G}}

\def\C{{\mathbb C}}

\newcommand{\tp}{\otimes}
\newcommand{\tc}{\!\circ\!}

\newcommand{\alg}[1]{\mathfrak{#1}}

\newcommand{\el}{\nonumber\\}

\def\ads{{AdS}_5\times {S}^5}

\def\bb{\mathbb}
\def\wt{\widetilde}
\def\wh{\widehat}
\def\ul{\underline}
\def\dg{\dagger}

\def\tE{{\widetilde E}}
\def\tF{{\widetilde F}}

\def\U{\mathcal{U}}

\def\psuc{\mathfrak{psu}\left(2|2\right)_{\mathcal{C}}}

\newcommand{\tl}[1]{\tilde{#1}}

\newcommand{\dwt}[1]{\widetilde{\widetilde{#1}}}
\newcommand{\dwh}[1]{\widehat{\widehat{#1}}}

\def\h{\mathfrak{a}}
\def\g{\mathfrak{g}}
\def\m{\mathfrak{b}}

\def\Yg{{\rm Y(\mathfrak{g})}}
\def\Ygh{{\rm Y(\mathfrak{g},\mathfrak{h})}}
\def\Ygg{{\rm Y(\mathfrak{g},\mathfrak{g})}}

\def\hh{\mathfrak{h}}
\def\mm{\mathfrak{m}}
\def\cc{\mathfrak{c}}

\def\afQ{\widehat{\cal{Q}}}
\def\afB{\widehat{\cal{B}}}

\newcommand{\rfl}{{\!re\!f\!}}

\title{Integrable boundaries in AdS/CFT: \\
revisiting the Z=0 giant graviton and D7-brane}

\author[a]{Marius de Leeuw}
\author[b,c]{and Vidas Regelskis}

\affiliation[a]{ETH Z\"urich, Institut f\"ur Theoretische Physik, \\Wolfgang-Pauli-Str.\ 27, CH-8093 Zurich, Switzerland}
\affiliation[b]{Department of Mathematics, University of York,\\Heslington, York YO10 5DD, UK}
\affiliation[c]{Institute of Theoretical Physics and Astronomy of Vilnius University,\\Go\v{s}tauto 12, Vilnius 01108, Lithuania}

\emailAdd{deleeuwm@phys.ethz.ch}
\emailAdd{vr509@york.ac.uk}

\abstract{We consider the worldsheet boundary scattering and the corresponding boundary
algebras for the $Z=0$ giant graviton and the $Z=0$ $D7$-brane in the AdS/CFT
correspondence. We consider two approaches to the boundary scattering, the usual
one governed by the (generalized) twisted Yangians and the $q$-deformed model of
these boundaries governed by the quantum affine coideal subalgebras. We show
that the $q$-deformed approach leads to boundary algebras that are of a
more compact form than the corresponding twisted Yangians, and thus are
favourable to use for explicit calculations. We obtain the $q$-deformed
reflection matrices for both boundaries which in the $q\to1$ limit specialize to
the ones obtained using twisted Yangians.}

\begin{document}

\maketitle


\section{Introduction}

The exploration of integrability in the AdS/CFT correspondence has revealed many different facets of integrability, that all have their origin in the planar limit of the duality between gauge and string theories (see review \cite{review}). One of the key directions of this exploration is the worldsheet scattering, which is largely driven by the centrally extended $\psuc$ algebra \cite{BeisertFundamental,BAnalytic,AFPZ} and its Yangian extension \cite{BeisertYangian}. These algebras play a central role in finding the relevant scattering matrices and writing the corresponding Bethe ansatz equations \cite{AFS,MdLrmat,AFBound,deLeeuw:2008ye,ALT}. This data is also of particular importance in solving the so-called $T$- and $Y$-systems used to describe the spectral problem \cite{Spectral} and calculating Wilson loops \cite{Wilson}.

A specific case of the worldsheet scattering is the boundary scattering which has attained lots of research interest and development on its own due to a large variety of the boundary conditions that arise when open strings end on $D$-branes embedded in the $\ads$ background (see \cite{McGreevy:2000cw,OpenB}). Boundary conditions depend not only on the type of the $D$-brane the string is attached to, but also on the type of embedding and the relative orientation of the string and the brane. The emerging integrable configurations have been classified in \cite{DOz}.

The best known and most studied boundaries are the so-called $Z=0$ and $Y=0$ giant gravitons that are $D3$-branes occupying the maximal $S^3\subset S^5$ of the $\ads$ spacetime \cite{D3refs,HM}, the $Z=0$ and $Y=0$ $D7$-branes wrapping $AdS_5 \times S^3$, and the ``horizontal'' and ``vertical'' $D5$-branes wrapping a defect hypersurface $AdS_4 \subset AdS_5$ and a maximal $S^2 \subset S^5$ \cite{D7D5refs,Kruczenski:2003be,CY,MR1}. While being six different configurations these actually provide only five different boundary conditions due to the equivalence of the $Y=0$ giant graviton and the $Y=0$ $D7$-brane from the worldsheet scattering point of view \cite{CY}.

The presence of boundaries generically breaks some of the underlying
symmetries. This makes the scattering more complicated than in the system without
the boundaries. Hence some more elaborate algebraic structures are needed to
solve the corresponding boundary scattering problem \cite{Sk}. The fundamental
AdS/CFT worldsheet $S$-matrix is determined up to an overall phase by the
underlying Lie algebra and the bound-state $S$-matrices are found by employing
the Yangian extension (see reviews \cite{LeeuwThesis}). However
finding boundary bound-state reflection matrices requires constructing coideal
subalgebras \cite{TwYangians}, e.g.\ the (generalized) twisted Yangians
\cite{DMS,MacKay:2002at,MK,VR}. These coideal subalgebras depend crucially on
the corresponding boundary conditions. The boundary scattering for the $Y=0$
giant graviton and the $Y=0$ $D7$-brane are identical and were shown to be
governed by a twisted Yangian of type I \cite{YangianY0,Palla1}, the boundary
scattering for the $Z=0$ giant graviton is governed by the twisted Yangian of
type II \cite{MR3}, and for the $D5$-brane it is the achiral twisted Yangian
\cite{MR4}. The boundary scattering for the $Z=0$ $D7$-brane is special as it
factorizes into non-equivalent left and right factors. The scattering in the
right factor is identical to the scattering for the $Z=0$ giant graviton, while
the left factor does not respect any supersymmetries and the boundary Yangian
structure for this case has not been revealed so far. The corresponding
reflection matrices have been found by solving the boundary Yang-Baxter equation
\cite{CY,MR1}. This is because the boundary Lie algebra alone is not enough to obtain even the fundamental reflection matrix. Thus knowing the boundary Yangian symmetry is necessary for algebraic Bethe ansatz techniques. Furthermore, this case is particularly interesting as it describes a non-supersymmetric boundary field theory. This shows that it is possible to break all supersymmetry without spoiling integrability, which manifests itself via symmetries of Yangian or quantum affine type.

In many cases, a Yangian algebra can be obtained as a specific ``degenerate'' limit of a quantum affine algebra. In the same way twisted Yangians are ``degenerate'' limits of twisted quantum affine coideal subalgebras. Quantum affine algebras, while being complex, are of a more elegant form than their Yangian counterparts. This becomes a very important feature when dealing with twisted Yangians.

A quantum affine algebra $\afQ$ leading to a $q$-deformed $S$-matrix which in
the $q\to1$ limit specializes to the AdS/CFT worldsheet $S$-matrix was
constructed in \cite{BGM} and the corresponding $q$-deformed bound-state
$S$-matrices were found in \cite{LMR}. However, finding fundamental scattering
matrices does not require the full quantum \textit{affine} algebra, thus the
fundamental $q$-deformed $S$-matrix was found earlier in \cite{BK}. This algebra
has also been employed in the Pohlmeyer reduced version of the AdS/CFT
superstring theory \cite{Hoare:2011fj}, which was shown to be important in
understanding the stability of the bound-states in the $q$-deformed theory
\cite{HoareBound}. 

The $q$-deformed boundary scattering for the $Y=0$ and $Z=0$ giant gravitons was first considered and the corresponding fundamental reflection matrices were obtained in \cite{MN}. A quantum affine coideal subalgebra governing boundary scattering for the $q$-deformed model of the $Y=0$ giant graviton was recently constructed in \cite{LMR2}. This coideal subalgebra was shown to be of a very compact form and may be considered as the quantum affine version of the quantum symmetric pairs developed in \cite{Le} (see also \cite{MRSHK}). In the rational $q\to1$ limit this algebra reproduces the twisted Yangian of the $Y=0$ giant graviton and the corresponding $q$-deformed bound-state reflection matrix specializes to the non-deformed one found in \cite{Palla1}. 

In the first part of this work we construct the twisted Yangian for the left factor of the $Z=0$ $D7$-brane, completing the quest of finding the boundary Yangians of the well known integrable boundaries in AdS/CFT. We also give a more elegant form of the Yangian symmetry of the $Z=0$ giant graviton found in \cite{MR3}. In the second part we construct quantum affine coideal subalgebras for $q$-deformed models of the $Z=0$ giant graviton and the left factor of the $Z=0$ $D7$-brane. These algebras have a rather compact form and follow the same pattern the one in \cite{LMR2}. The compactness of the algebra is very important for the $Z=0$ giant graviton since the twisted Yangian of it is of a complicated form, thus in this case it is much more convenient to deal with the quantum affine coideal subalgebra than with the corresponding twisted Yangian. The \mbox{$q$-deformed} reflection matrices of these models in the $q\to1$ limit specialize to the ones found in \cite{HM,CY,MR1}.     

This paper is organized as follows. In section \ref{sec:2} we review the worldsheet scattering and the boundary scattering for the $Z=0$ giant graviton and the $D7$-brane, and the corresponding boundary symmetries. In section \ref{sec:3} we recall the AdS/CFT Yangian symmetry and the twisted Yangian of the $Z=0$ giant graviton, and construct the twisted Yangian for the left factor of the $D7$-brane. In section \ref{sec:4} we construct the quantum affine coideal subalgebras for the $q$-deformed models of the $Z=0$ giant graviton and the left factor of the $D7$-brane. Section \ref{sec:5} contains discussion and concluding remarks. The $q$-deformed reflection matrices of the $Z=0$ giant graviton are given in the Appendix \ref{AppA}. 


\section{The setup}\label{sec:2}


In this section we will first recall the symmetry properties and the necessary preliminaries of the worldsheet scattering and reflection matrices. After this we continue with a short discussion on the two different boundary problems that we address in this paper.


\subsection{Scattering and reflection}


\paragraph{The algebra.} 

The symmetry algebra of excitations in the light-cone string theory on the $\ads$ background and for the single-trace local operators in the $\mathcal{N}=4$ supersymmetric Yang-Mills gauge theory is given by two copies (left and right) of the centrally-extended Lie superalgebra \cite{BeisertFundamental,AFPZ}
\begin{equation}
\psuc = \mathfrak{psu}\left(2|2\right)\ltimes\bb{R}^{3}\,.\label{g}
\end{equation}
This Lie algebra contains two sets of bosonic $\mathfrak{su}(2)$ rotation generators $\bb{R}_{a}^{\; b}$, $\bb{L}_{\alpha}^{\; \beta}$, two sets of fermionic supersymmetry generators $\bb{Q}_{\alpha}^{\; a},$ $\bb{G}_{a}^{\; \alpha}$ and three central charges $\bb{C}$, $\bb{C}^\dg$ and $\bb{H}$. The non-trivial commutation relations are
\begin{align}
{[\,\bb{L}_{\alpha}^{\;\beta},\bb{J}_{\gamma}]} &=
\delta_{\gamma}^{\beta}\bb{J}_{\alpha}-\tfrac{1}{2}\delta_{\alpha}^{\beta}\bb{J}
_{\gamma}\,,
&\{ \bb{Q}_{\alpha}^{\; a},\bb{Q}_{\beta}^{\; b} \} &=
\epsilon^{ab}\epsilon_{\alpha\beta}\bb{C}\,, \el
{[\,\bb{L}_{\alpha}^{\; \beta},\bb{J}^{\gamma}]}
&=-\delta_{\alpha}^{\gamma}\bb{J}^{\beta}+\tfrac{1}{2}\delta_{\alpha}^{\beta}\bb
{J}^{\gamma}\,,
&\{ \bb{G}_{a}^{\; \alpha},\bb{G}_{b}^{\;
\beta}\}&=\epsilon^{\alpha\beta}\epsilon_{ab}\bb{C}^{\dg}\,, \el
{[\,\bb{R}_{a}^{\; b},\bb{J}_{c}]} &=
\delta_{c}^{b}\bb{J}_{a}-\tfrac{1}{2}\delta_{a}^{b}\bb{J}_{c}\,,
&\{ \bb{Q}_{\alpha}^{\; a},\bb{G}_{b}^{\;
\beta}\}&=\delta_{b}^{a}\bb{L}_{\beta}^{\;
\alpha}+\delta_{\beta}^{\alpha}\bb{R}_{b}^{\;
a}+\tfrac{1}{2}\delta_{b}^{a}\delta_{\beta}^{\alpha}\bb{H}\,, \el
{[\,\bb{R}_{a}^{\; b},\bb{J}^{c}]} &=
-\delta_{a}^{c}\bb{J}^{b}+\tfrac{1}{2}\delta_{a}^{b}\bb{J}^{c}\,,
\label{Lie_algebra}
\end{align}
where $a,\; b,...=1,\,2$ and $\alpha,\;\beta,...=3,\,4$, and the symbols $\bb{J}_{a}$, $\bb{J}_{\alpha}$ with lower (or upper) indices represent any generators with the corresponding index structure.

This algebra may be equipped with a non-trivial (braided) Hopf algebra structure \cite{Gomez:2006va} such that for any $\bb{J}^A \in \psuc$ 
\begin{equation} \label{coprodot}
\Delta (\bb{J}^A) = \bb{J}^A \otimes 1 + \U^{[[A]]} \otimes \bb{J}^A \,, \qquad\quad
\Delta^{op} (\bb{J}^A) = \bb{J}^A \otimes \U^{[[A]]} + 1 \otimes \bb{J}^A \,.
\end{equation}
Here $\U$ is the so-called braiding factor of the algebra and the additive quantum number $[[A]]$ equals $0$ for generators in $\alg{sl}(2)\oplus \alg{sl}(2)$ and for $\bb{H}$, $\frac{1}{2}$ for $\bb{Q}_{\alpha}^{\ a}$, $-\frac{1}{2}$ for $\bb{G}^{\alpha}_{a}$, $1$ for $\bb{C}$ and $-1$ for $\bb{C}^\dag$.


\paragraph{Representations.} 

The physical excitations of the $\ads$ superstring transform in the supersymmetric short representations of $\psuc$ \cite{BAnalytic}. They are conveniently described in terms of the superspace formalism introduced in \cite{AFBound}. The representation describing an \mbox{$M$-particle} bound-states consists of vectors $|m,n,k,l\rangle\in V(p)$ where $k+l+m+n=M$ and $V(p)$ is the corresponding vector space of excitations with momentum $p$. The labels $m,n$ denote fermionic degrees of freedom and $k,l$ denote the bosonic part. The symmetry generators act on the basis vectors as
\begin{align}
 & \bb{R}^{\;1}_2\,|m,n,k,l\rangle = k \,|m,n,k-1,l+1\rangle\,, &  & \bb{L}^{\;4}_3\, |m,n,k,l\rangle = |m+1,n-1,k,l\rangle\,,\el
 & \bb{R}^{\;2}_1\,|m,n,k,l\rangle = l \,|m,n,k+1,l-1\rangle\,, &  & \bb{L}^{\;3}_4\, |m,n,k,l\rangle = |m-1,n+1,k,l\rangle\,.
\end{align}
The action of the supercharges is given by
\begin{align}
\bb{Q}^{\;2}_4\, |m,n,k,l\rangle = & ~a~(-1)^{m} l \,|m,n+1,k,l-1\rangle + b~|m-1,n,k+1,l\rangle\,,\el
\bb{G}^{\;4}_2\, |m,n,k,l\rangle = & ~c~k \,|m+1,n,k-1,l\rangle+d~(-1)^{m}\,|m,n-1,k,l+1\rangle\,.
\end{align}
The explicit action of the rest of the charges is easily obtained by employing the commutation relations \eqref{Lie_algebra}.  A convenient parametrization of the representation labels of the states in the bulk is \cite{BeisertFundamental,AFBound}
\begin{align}
a=\sqrt{\frac{g}{M}}\gamma,\quad
b=\sqrt{\frac{g}{M}}\frac{\alpha}{\gamma}\left(1-\frac{x^{+}}{x^{-}}\right),
\quad 
c=\sqrt{\frac{g}{M}}\frac{i\gamma}{\alpha x^{+}},\quad
d=\sqrt{\frac{g}{M}}\frac{i x^{+}}{\gamma}\left(\frac{x^{-}}{x^{+}}-1\right),
\label{abcd}
\end{align}
where $M$ is the bound-state number ($M=1$ corresponds to the fundamental representation), $g$ is the coupling constant, and $x^{\pm}$ are the spectral parameters ($e^{ip}=\frac{x^{+}}{x^{-}}$) respecting the mass-shell (multiplet--shortening) constraint
\begin{equation}
x^{+}\!+\frac{1}{x^{+}}-x^{-}\!-\frac{1}{x^{-}}=\frac{iM}{g}\,.\label{shortening}
\end{equation}
The parameters $\gamma$ and $\alpha$ are internal parameters of the representation and define the relative normalization between bosons and fermions. The unitarity requirement imposes $\alpha^\dag=\alpha^{-1}$ and $\gamma={\rm e}^{i{\varphi}}\sqrt{i\left(x^{-}\!-x^{+}\right)}$~, where the arbitrary phase factor ${\rm e}^{i\varphi}$ reflects the freedom in choosing $x^{\pm}$. The rapidity of the magnon in the $x^\pm$ parametrization is defined to be
\begin{equation}
u=x^{+}\!+\frac{1}{x^{+}} - \frac{i M}{2g}\,.\label{rapidity}
\end{equation}
The eigenvalues of the central charges for the $M$-magnon bound-states are expressed by
\begin{eqnarray}
 & C_{M} = M\, ab = g\alpha\left(1-{\rm e}^{ip}\right), \qquad 
C_{M}^{\dg} = M\, cd = g\alpha^{-1}\left(1-{\rm e}^{-ip}\right) ,\el 
 & H_{M}=M\left(ad+bc\right)=\sqrt{M^{2}+16g^{2}\sin^{2}\frac{p}{2}} \,. \label{CCH}
\end{eqnarray}
%


\paragraph{Bulk scattering theory.}

The fundamental scattering matrix 
\begin{equation}
S:V(p_1)\otimes V(p_2) \longrightarrow V(p_1)\otimes V(p_2)\,,
\end{equation}
is obtained by requiring it to respect the symmetry algebra, i.e.\ to intertwine the coproduct and the opposite coproduct
\begin{equation} \label{invar}
\Delta^{op}(\bb{J}^A) \, S = S \, \Delta(\bb{J}^A) ~.
\end{equation}
The above requirement is restrictive enough to fix the matrix structure for the
fundamental $S$-matrix. However, the Lie algebra alone is not enough to define all
coefficients of the generic bound-state $S$-matrices uniquely. This is because the tensor product of the higher order supersymmetric short representations generically yields a sum of irreducible long representations. To remedy this, one either needs to invoke the Yang-Baxter equation or use Yangian symmetry \cite{MdLrmat,AFBound}. 


\paragraph{Reflection algebra and the boundary scattering.} 

The reflection matrix maps incoming states with momentum $p$ to outgoing states with momentum $-p$ while keeping the boundary states invariant under the reflection,
\begin{equation} \label{eq:reflection}
K : V(p) \otimes V(s) \longrightarrow V(-p) \otimes V(s) \,.
\end{equation}
Here $V(p)$ represents the vector space of the bulk states and $V(s)$ represents the vector space of the boundary states with $s$ denoting any parameters associated to the boundary states. 

The representation labels associated to the reflected states in $V(-p)$ can be obtained from \eqref{abcd} using the reflection map $\kappa : x^\pm \mapsto - x^\mp$ leading to a matrix relation between the representation labels of incoming and outgoing states,
\begin{equation}
\begin{pmatrix}\underline{a}&\underline{b}\\ \underline{c}&\underline{d}
\end{pmatrix} D
=T\begin{pmatrix}a&b\\ c&d \end{pmatrix}T^{-1} 
\quad\text{with}\quad
D=\begin{pmatrix} \gamma/\ul{\gamma} & 0 \\ 0 & \ul{\gamma}/\gamma
\end{pmatrix}, 
\quad
T=\begin{pmatrix} \U^{-2} & 0 \\ 0 & -1 \end{pmatrix}, \label{abcd1_ref}
\end{equation}
where the underbarred parameters are the image of the usual representation paremeters under the reflection map, i.e.\ $
\kappa : (a,b,c,d) \mapsto (\ul{a},\ul{b},\ul{c},\ul{d})$ and $\kappa : \gamma \mapsto \ul{\gamma}$. This notation allows us to introduce the reflected coproduct \cite{MR3},
\begin{equation} \label{cop_ref}
\Delta^{ref} (\bb{J}^A) = \ul{\bb{J}}^A \otimes 1 + \U^{-[[A]]} \otimes {\bb{J}}^A \,,
\end{equation}
where $\Delta^{ref} = (\kappa \otimes 1 ) \circ \Delta$ and $\ul{\bb{J}}^A = \kappa(\bb{J}^A)$, and $\kappa(\U)=\U^{-1}$ has been used implicitly.%
\footnote{This construction is algebra specific, because the reflection map $\kappa: \U \to \U^{-1}$ is an involution of the algebra and leads to the representations of $\psuc$ for incoming and reflected states. However a reflection map for an arbitrary Lie algebra can be explicitly constructed at the representation level only.} 

Suppose the  boundary preserves a subalgebra $\h$ of $\psuc$. Then we can formulate the symmetry properties of the reflection matrix $K$ in a similar way as those of the $S$-matrix \eqref{invar}. Namely, the symmetry properties of the reflection matrix are simply given by the boundary intertwining equation
\begin{equation} \label{eq:int_B}
\Delta^{ref} (\bb{J}^A)\, K = K \,\Delta(\bb{J}^A) \,,
\end{equation}
with $\bb{J}^A \in\h$, and the reflection matrix is required to satisfy the reflection (the boundary Yang-Baxter) equation
\begin{equation}
\K_{2} S_{2\underline{1}}\K_{1} S_{12} = S_{\underline{2}\underline{1}}\K_{1} S_{1\underline{2}}\K_{2}\,.
\end{equation}
Here the underbarred notation denotes reflected states. 

We will now proceed with a discussion of the two different types of boundaries that we consider in this paper.


\subsection{The \texorpdfstring{$Z=0$}{Z=0} giant graviton}

The maximal giant graviton is a $D3$-brane in the $AdS_{5}\times S^{5}$ spacetime wrapping a topologi\-cally-trivial cycle enclosing maximal $S^{3}\subset S^{5}$, and is prevented from collapsing by coupling to the background supergravity fields \cite{McGreevy:2000cw}. The usual parametrization of $S^{5}$ is expressed in terms of the complex coordinates $X=\Phi_{1}+i\Phi_{2}$, $Y=\Phi_{3}+i\Phi_{4}$, $Z=\Phi_{5}+i\Phi_{6}$ respecting $|X|^{2}+|Y|^{2}+|Z|^{2}=1$, where the radius of $S^{5}$ has been set to unity, $R=1$. In this parametrization the maximal giant graviton is obtained by setting any two $\Phi_{i}$'s to zero. However, any two such configurations are related to each other by an $SO(6)$ rotation. This symmetry can be broken by attaching an open string to the brane and giving it a charge $J$ corresponding to the preferred $SO(2)\subset SO(6)$ rotation. 

The parametrization in complex coordinates makes it easy to translate this setup
to the gauge theory side. The triplet $X$, $Y$, $Z$ can be thought of as
representing the three complex scalar fields of the ${\cal N}=4$ super
Yang-Mills. Then the field theory description of the string in the large $J$
limit carries a large number of insertions, called the Bethe vacuum state, of
the field corresponding to the preferred rotation, and a relatively small number
of other fields, called excitations (or simply magnons). The explicit
description of the string in the gauge theory depends on the choice of the
particular generator $J$ and the relevant orientation of the giant graviton
inside $S^{5}$. The two relevant cases are obtained by choosing $J=J_{56}$ and
the giant graviton to be the maximal three sphere given by $Y=0$ or $Z=0$ with
the standard Bethe vacuum on the string being $Z=X_{5}+iX_{6}$ \cite{HM}. 

In the large $J$ limit the string worldsheet is a very long segment.
Consequently, the left and right boundaries are well separated and can be
treated independently; thus the boundary scattering becomes equivalent to
scattering on a semi-infinite line. In the AdS/CFT this translates into the
description of a magnon incoming from infinity, reflecting at the boundary,
and returning back to infinity. Hence the asymptotic states are interpolating
between the usual vacuum of BMN states \cite{BMN} and the boundary. This
treatments allows us to employ the usual $S$-matrix technique discussed above to study the
boundary scattering. 

Here we will consider the $Z=0$ giant graviton which preserves the same supersymmetries as the field $Z$, and thus the boundary Lie algebra is $\psuc$.


\paragraph{Boundary representation.}

The boundary forms a supersymmetric short representation of the Lie algebra $\psuc$. This representation is parametrized by the following labels \cite{HM}
\begin{equation}
a_{B}=\sqrt{\frac{g}{M}}\gamma_{B},\quad
b_{B}=\sqrt{\frac{g}{M}}\frac{\alpha}{\gamma_{B}},\quad
c_{B}=\sqrt{\frac{g}{M}}\frac{i\gamma_{B}}{\alpha x_{B}},\quad
d_{B}=\sqrt{\frac{g}{M}}\frac{x_{B}}{i\gamma_{B}}\,, \label{abcd_B}
\end{equation}
and the multiplet shortening (mass-shell) condition is
\begin{equation} \label{shortening_B}
x_{B}+\frac{1}{x_{B}}=\frac{iM}{g} \,.
\end{equation}
The boundary values of the central charges are
\begin{eqnarray}
 & C_{(B)M} = M\, a_B b_B = g\alpha\,, \qquad 
C_{(B)M}^{\dg} = M\, c_B d_B = g\alpha^{-1}\,,\el 
 & H_{(B)M}=M\left(a_B d_B + b_B c_B \right)=\sqrt{M^{2}+4g^{2}} \,. \label{CCH_B}
\end{eqnarray}
The unitarity requirement imposes an additional constraint, $\gamma_B = e^{i \varphi_B}\sqrt{-i x_B}$. Thus this representation is just an $M$-particle bound-state representation with different labels. Interestingly, boundary labels can be obtained from the bulk ones in \eqref{abcd} by a simple bulk-to-boundary map $x^\pm \mapsto \pm x_B$ together with a rescaling of the coupling constant $g\to g/2$. This rescaling is introduced to cancel the factor of $\sqrt{2}$ appearing due to the bulk-to-boundary map of $\gamma$, i.e.\ $\gamma \mapsto \sqrt{2} \, \gamma_B$. This map also reproduces \eqref{CCH_B} when applied to \eqref{CCH}. In such a way the $M$-magnon boundary bound-state can be interpreted as a bulk $2M$-magnon bound-state with a maximal momentum, $p=\pi$, i.e.\ it is the state at the end of the Brillouin zone. 

Finally, note that the braiding factor $\U$, which is a central (and group-like) element of the $\psuc$ algebra, is not in the boundary algebra, and thus, strictly speaking, the boundary algebra is a subalgebra of the bulk algebra isomorphic to $\psuc$, but parametrized by one parameter -- the coupling constant $g$ only.%
\footnote{The parameter $\alpha$ can be neglected.}
Hence the boundary algebra is invariant under the reflection map $\kappa$.


\paragraph{Scattering.} 

Reflection matrix for the fundamental particles was found in \cite{HM} by using the boundary Lie algebra, and the reflection matrices for the 2-particle bound-states were found in \cite{MR1} by using boundary Lie algebra combined with the reflection equation. These reflection matrices were shown to follow from the twisted Yangian structure \cite{MR3}. In later sections we give a more elegant form this symmetry.


\subsection{The \texorpdfstring{$Z=0$}{Z=0} \texorpdfstring{$D7$}{D7}-brane}

The second system we will consider is the so-called ``$Z=0$ $D7$-brane'' configuration, where the $D7$-brane is wrapping the entire $AdS_{5}$ and a maximal $S^{3}\subset S^{5}$ of the underlying $\ads$ background. Boundary scattering for this system was presented in \cite{CY}. Here we will briefly reall the properties of this configuration that are relevant to us.

The $Z=0$ $D7$-brane is obtained by setting $X_{5}=X_{6}=0$ in the parametrization of $S^5$. This choice breaks the $SO(6)$ symmetry down to $SO(4)_{1234}\times SO(2)_{56}$. It also breaks exactly half of the background supersymmetries that are left handed with respect to the boundary $SO(4)$ symmetry, and the surviving fields on the gauge theory side form the $\mathcal{N}=2$ chiral hypermultiplet \cite{Kruczenski:2003be}. Next, choosing the Bethe vacuum to be $Z=X_{5}+iX_{6}$ and the preferred charge $J=J_{56}$ rotating the directions transverse to the brane thus preserving the $SO(4)$ symmetry, one further breaks half of the residual supercharges -- the left copy of $\psuc$. This leaves the boundary algebra to be 
\begin{equation}
\mathfrak{su}(2)\times\mathfrak{su}(2)\times\widetilde{\mathfrak{psu}}(2|2)\ltimes\mathbb{R}^{3}\,.
\end{equation}
The fundamental matter fields transform in a $(1,\boxslash)$ representation of $\mathfrak{psu}(2|2)\times\widetilde{\mathfrak{psu}} (2|2)$ (we refer to \cite{CY} for the explicit details on the boundary matter content). 

This setup leads to a factorization $K\otimes\wt{K}$ of the complete reflection matrix, and thus two independent reflection processes need to be considered, the reflection in the left and in the right factor of the brane. 

The reflection in the right factor
\begin{equation}
\wt{K} : V(p)\otimes V(s) \longrightarrow V(-p)\otimes V(s) \,,
\end{equation}
is equivalent to the reflection from the $Z=0$ giant graviton discussed above. The reflection in the left factor 
\begin{equation}
K : V(p)\otimes 1 \longrightarrow V(-p)\otimes 1\,.
\end{equation}
is a reflection from a non-supersymmetric singlet boundary. The fundamental reflection matrix was found in \cite{CY}, the bound-state one was found in \cite{MR1}. Let us now recall the latter case.


\paragraph{Scattering theory.}

The boundary we are considering is a singlet with respect to the boundary algebra, thus it may be represented by the boundary vacuum state $|0\rangle_{B}$ which is annihilated by all generators of the boundary algebra \cite{GZ}.
We define the reflection matrix to be the intertwining matrix 
\begin{equation}
K\,|m,n,k,l\rangle\otimes|0\rangle_{B}=K_{(m,n,k,l)}^{(a,b,c,d)}\,|a,b,c,
d\rangle\otimes|0\rangle_{B} \,.\label{KX}
\end{equation}
The space of states $|m,n,k,l\rangle$ is $4M$-dimensional and can be decomposed into four $4M=(M+1)+(M-1)+M+M$ subspaces that have the orthogonal basis 
\begin{align}
|k\rangle^{1} & =|0,0,k,M\!-\! k\rangle \,, & & k=0\ldots M \,,\el
|k\rangle^{2} & =|1,1,k\!-\!1,M\!-\! k\!-\!1\rangle \,, & & k=1\ldots M-1 \,,\el
|k\rangle^{3} & =|1,0,k,M\!-\! k\!-\!1\rangle \,, & & k=0\ldots M-1 \,,\el
|k\rangle^{4} & =|0,1,k,M\!-\! k\!-\!1\rangle \,, & & k=0\ldots M-1 \,.
\label{KXbasis}
\end{align}

The boundary Lie algebra in the left factor is generated by the bosonic generators $\R_a^{\; b}$ and $\L_\alpha^{\; \beta}$, and the central charge $\bb{H}$ only. It constrains the reflection matrix $K$ to be diagonal for any $k$ and $M$,
\begin{equation}
K\,|k\rangle^{1} = A \,|k\rangle^{1}\,, \quad\qquad K\,|k\rangle^{2} = B
\,|k\rangle^{2}\,, \quad\qquad K\,|k\rangle^{\alpha} = C \,|k\rangle^{\alpha}
\,, \label{KXref}
\end{equation}
where $\alpha=3,\,4$ and we have dropped the boundary vacuum state. The standard
normalization is $A=1$. This leaves coefficients $B$ and $C$ undetermined.
However, due to a simple form of the reflection matrix, these can
readily be found by solving the boundary Yang-Baxter equation. It
factorizes in this case, and thus can be solved by the method of separating
variables. Consequently one finds
\begin{equation}
B =\frac{x_B+x^+}{x_B-x^-}\frac{\ul{\gamma}}{\gamma} \,, \qquad\qquad 
C =\frac{(x_{B}+x^{+})(1-x_{B}x^{+})}{(x_{B}-x^{-})(1+x_{B}x^{-})}\frac{\ul{\gamma}^{2}}{\gamma^{2}} \,, \label{XBC}
\end{equation}
where the parameter $x_B$ satisfies the fundamental mass-shell condition $x_B +
1/x_B = i /g$. This constraint is obtained by considering the ``supersymmetric''
matrix elements of the boundary Yang-Baxter equation, e.g.\ ${}^3\langle k_i |
\otimes {}^4\langle k_j | \,{\rm BYBE}\, | k_m \rangle^1 \otimes | k_n
\rangle^1$. We refer to \cite{CY,MR1} for details.

In the next section we will construct the boundary Yangian algebra and show that it leads to the same reflection matrix.


\section{Boundary Yangian algebras}\label{sec:3}


The crucial algebraic structure that allows us to fully determine the matrix structure of the AdS/CFT scattering matrices is the Yangian of $\psuc$. For reflection from a boundary the analogous structure is that of a twisted Yangian. In this section we briefly discuss this algebraic framework and then specialize it to the two boundary models we have discussed in the previous section.


\subsection{Yangians and reflection}


\paragraph{Yangian $\Yg$.} \label{sec:3.1}

The Yangian $\Yg$ of a Lie algebra $\g$ is a deformation of the universal enveloping algebra of the polynomial algebra $\g[u]$. It has level-0 $\g$ generators $\bb{J}^{a}$ and level-1 $\mbox{Y}(\g)$ generators $\wh{\bb{J}}^{a}$. Their commutators have the generic form
\begin{equation}
[\,\bb{J}^{a},\bb{J}^{b}] = f_{\quad\! c}^{ab}\,\bb{J}^{c}, 
\qquad [\,\bb{J}^{a},\wh{\bb{J}}^{b}] = f_{\quad\! c}^{ab}\,\wh{\bb{J}}^{c},
\end{equation}
and are required to obey Jacobi and Serre relations \cite{MK}
\begin{equation}
\bigl[\bb{J}^{[a},\bigl[\,\bb{J}^b,\bb{J}^{c]}\bigr]\bigr] = 0\,, \qquad\quad \bigl[\,\wh{\bb{J}}^{[a},\bigl[\,\wh{\bb{J}}^b,\bb{J}^{c]}\bigr]\bigr] = 
\mathcal{O}(\bb{J}^3)\,, \label{Serre}
\end{equation}
where $^{[a\,b\,c]}$ denotes cyclic permutations, and indices $a\,(,b,c,...)$ run over all generators of $\g$. Indices of the structure constants $f_{\quad\!d}^{ab}$ can be lowered
by means of the inverse Killing--Cartan form.
The co-product of the generators then takes the form
\begin{equation} \label{eqn;coprodYangian}
\Delta(\bb{J}^{a}) = \bb{J}^{a}\otimes1+1\otimes\bb{J}^{a},
\qquad\Delta(\wh{\bb{J}}^{a}) = \wh{\bb{J}}^{a}\otimes1+1\otimes\wh{\bb{J}}^{a}+\tfrac{1}{2}f_{\;
bc}^{a}\,\bb{J}^{b}\otimes\bb{J}^{c}.
\end{equation}
Finite-dimensional representations of $\mbox{Y}(\g)$ are realized in one-parameter families, via the evaluation automorphism
\begin{equation} \label{one_param}
\tau_{v}:\mbox{Y}(\g)\rightarrow\mbox{Y}(\g)\,,\quad\bb{J}^{a}\mapsto\bb{J}^{a}\,,
\quad\wh{\bb{J}}^{a}\mapsto\wh{\bb{J}}^{a}+v\,\bb{J}^{a}\,,
\end{equation}
corresponding to a shift in the polynomial variable. Some finite-dimensional irreducible representations of $\g$ may be extended to representations of $\mbox{Y}(\g)$ via the evaluation map
\begin{equation}
\mathrm{ev}_{v}:\mbox{Y}(\g)\rightarrow\mbox{U}(\g)\,,\quad\bb{J}^{a}\mapsto\bb{J}^{a}\,,\quad\wh{\bb{J}}^{a}\mapsto v\,\bb{J}^{a}\,,\label{ev_map}
\end{equation}
which yields evaluation modules, with states $\left|v\right\rangle $ carrying a spectral parameter $v$.

We will build finite-dimensional representations of $\mbox{Y}(\g)$ by considering the tensor product of two such $\g$-modules on which the bulk $S$-matrix acts. The action of Yangian generators on the $\g$-module $V(v)$ is then defined correspondingly
\begin{equation}
\wh{\bb{J}}^a\left|v\right\rangle =\gamma\,(v+v_{0})\,\bb{J}^a\left|v\right\rangle
,\qquad\left|v\right\rangle \in V\left(v\right),\label{vJ_ansatz}
\end{equation}
with $\gamma$ some $\bb{C}$-number to be determined and $v_{0}$ some representation parameter.


\paragraph{Twisted Yangian $\mbox{Y}(\g,\h)$.}

Consider an integrable model with the symmetry algebra given by the Yangian
$\mbox{Y}(\g)$ of some Lie algebra $\g$. Suppose that the boundary module
respects a subalgebra $\h\subset\g$ corresponding to an involution
$\theta:\g\rightarrow\g$ such that $\h$ is left invariant under $\theta$, {\it i.e.} $\h=\g^\theta$. Then $\h$ and the subset
$\m=\g \backslash \h$ respecting
\begin{equation} \label{symmetric_pair}
\left[\h,\h\right]\subset\h,\qquad\left[\h,\m\right]\subset\m,\qquad\left[\m,
\m\right]\subset\h \,.
\end{equation}
are the positive and negative eigenspaces of $\theta$, namely $\theta(\h)=+\h$ and $\theta(\m)=-\m$. The associated symmetry algebra respected by the boundary is the so-called (generalized) twisted Yangian $Y(\g,\h)$ of type I \cite{DMS} (see also \cite{MacKay:2002at,VR}) generated by the level-0 charges $\mathbb{J}^{i}$ and twisted level-1 charges
\begin{equation} \label{twist}
\wt{\bb{J}}^{p} := \wh{\bb{J}}^{p} + t \, \bb{J}^{p} + \tfrac{1}{4}f_{\;\;
qi}^{p}\left(\bb{J}^{q}\,\bb{J}^{i}+\bb{J}^{i}\,\bb{J}^{q}\right) , 
\end{equation}
where indices $i(,j,k,...)$ run over the $\h$-indices and $p,q(,r,...)$ over the $\m$-indices. The parameter $t$ corresponds to the freedom of shifting the spectral parameter via the automorphism of the Yangian \eqref{one_param}. It can be restricted to a particular value by some additional constraints of the algebra or by solving the boundary intertwining equation. The coproducts of the charges $\bb{J}^{i}$ and $\wt{\bb{J}}^{i}$ are
\begin{equation} \label{twistcop}
\Delta(\bb{J}^{i}) = \bb{J}^{i}\otimes1+1\otimes\bb{J}^{i} \,,
\qquad\Delta(\wt{\bb{J}}^{p}) = \wt{\bb{J}}^{p}\otimes1 + 1\otimes\wt{\bb{J}}^{p} +
f_{\;\; qi}^{p}\,\bb{J}^{q}\otimes\bb{J}^{i} \,,
\end{equation}
and satisfy the co-ideal property
\begin{equation} \label{coideal_g_h}
\Delta\mbox{Y}(\g,\h)\subset\mbox{Y}(\g)\otimes\mbox{Y}(\g,\h) \,.
\end{equation}

This construction is not valid when $\theta$ is trivial, i.e.\ $\g^\theta=\g$. This case corresponds to the twisted Yangian $\mbox{Y}(\g,\g)$ described below. 


\paragraph{Twisted Yangian $\mbox{Y}(\g,\g)$.} 

Consider a boundary which respects all of the bulk Lie algebra $\g$. Such a boundary does not respect any level-1 charges and the corresponding boundary Yangian is a twisted Yangian $\Ygg$ of type II generated by level-0 generators $\bb{J}^a$ and twisted level-2 charges \cite{VR}
\begin{equation} \label{J2tw}
\dwt{\bb{J}}{}^{a} = \wh{\wh{\bb{J}}}{}^{a} + t \, \wh{\bb{J}}^a + \frac{1}{4}f^{a}_{\; bc}\big(\,\wh{\bb{J}}^{b}\bb{J}^{c}+\bb{J}^{c}\wh{\bb{J}}^{b}\big) \,,
\qquad \dwh{\bb{J}}{}^{a} = \frac{1}{c_{\g}}f^{a}_{\;\,bc}\,[\,\wh{\bb{J}}^{c},\wh{\bb{J}}^b] \,,
\end{equation}
having coproducts of the form
\begin{equation} \label{J2twcop}
\Delta(\dwt{\bb{J}}{}^{a}) = \dwt{\bb{J}}{}^{a}\otimes1+1\otimes\dwt{\bb{J}}{}^{a} + f^a_{\; bc}\,\wh{\bb{J}}^{b}\otimes \bb{J}^{c} + \frac{1}{4c_\g}f_{\; bc}^{a} \big(h^{\;\; cb}_{+\; lki}\, \bb{J}^{l}\bb{J}^{k}\otimes \bb{J}^{i} + h^{\;\; cb}_{-\; lki}\, \bb{J}^{i}\otimes \bb{J}^{l}\bb{J}^{k} \big) \,,
\end{equation}
and satisfying the coideal property
\begin{equation} \label{
coideal_g_g}
\Delta\mbox{Y}(\g,\g)\subset\mbox{Y}(\g)\otimes\mbox{Y}(\g,\g) \,,
\end{equation}
where $h^{\;\; cb}_{\pm\; lki} = f_{\; ld}^{c}f_{\; ke}^{b}f_{\quad\! i}^{de} \pm f_{\quad\! d}^{ce}(f_{\; ke}^{b}f_{\; li}^{d} + f_{\; le}^{b}f_{\; ki}^{d})$ and $c_\g$ is the eigenvalue of the quadratic Casimir operator in the adjoint representation. Indices $a(,b,c,...)$ run over all indices of $\g$, and $t$ is an arbitrary complex parameter playing the same role as in \eqref{twist}, i.e.\ can be restricted to a particular value by some additional constraints.


\paragraph{Yangian ${\rm Y(\psuc)}$.} 

The Yangian symmetry of the worldsheet $S$-matrix is generated by the Lie algebra \eqref{g} and the corresponding Yangian generators having the following coproducts \cite{BeisertYangian}
\begin{align}
\Delta(\wh{\bb{R}}_{a}^{\; b}) \, &= \wh{\bb{R}}_{a}^{\;b}\otimes1+1\otimes\wh{\bb{R}}_{a}^{\; b}+\tfrac{1}{2}\bb{R}_{a}^{\;c}\otimes\bb{R}_{c}^{\; b}-\tfrac{1}{2}\bb{R}_{c}^{\; b}\otimes\bb{R}_{a}^{\; c}
- \tfrac{1}{2}\bb{G}_{a}^{\;\gamma}\,\mathcal{U}^{+1}\!\otimes\bb{Q}_{\gamma}^{\; b}\el
& \qquad -\tfrac{1}{2}\bb{Q}_{\gamma}^{\;b}\,\mathcal{U}^{-1}\!\otimes\bb{G}_{a}^{\; \gamma}
+\tfrac{1}{4}\delta_{a}^{b}\bb{G}_{c}^{\;\gamma}\,\mathcal{U}^{+1}\!\otimes\bb{Q}_{\gamma}^{\;
c}+\tfrac{1}{4}\delta_{a}^{b}\bb{Q}_{\gamma}^{\;c}\,\mathcal{U}^{-1}\!\otimes\bb{G}_{c}^{\; \gamma}\,, \el
\Delta(\wh{\bb{L}}_{\alpha}^{\enskip\beta}) &= \wh{\bb{L}}_{\alpha}^{\enskip\beta} \otimes 1+1 \otimes \wh{\bb{L}}_{\alpha}^{\enskip\beta}-\tfrac{1}{2}\,\bb{L}_{\alpha}^{\enskip\gamma} \otimes \bb{L}_{\gamma}^{\enskip\beta}+\tfrac{1}{2}\,\bb{L}_{\gamma}^{\enskip\beta} \otimes \bb{L}_{\alpha}^{\enskip\gamma} +\tfrac{1}{2}\,\mathcal{U}^{+1}\,\bb{G}_{c}^{\enskip\beta} \otimes \bb{Q}_{\alpha}^{\enskip c} \el
 & \qquad +\tfrac{1}{2}\,\mathcal{U}^{-1}\,\bb{Q}_{\alpha}^{\enskip c} \otimes \bb{G}_{c}^{\enskip\beta}-\tfrac{1}{4}\,\delta_{\alpha}^{\beta}\,\mathcal{U}^{+1}\,\bb{G}_{c}^{\enskip\gamma} \otimes \bb{Q}_{\gamma}^{\enskip c}-\tfrac{1}{4}\,\delta_{\alpha}^{\beta}\,\mathcal{U}^{-1}\,\bb{Q}_{\gamma}^{\enskip c}\, \otimes \bb{G}_{c}^{\enskip\gamma} \,, \label{Y(g)}
\end{align}
and the rest can be obtained by means of the commutation relations \eqref{Lie_algebra}. These Yangian generators can be written in a very elegant form by employing the Casimir-like operator
\begin{align} \label{CasT}
\bb{T} = \bb{R}_a^{\;b}\,\bb{R}_b^{\;a} - \bb{L}_\alpha^{\;\beta}\,\bb{L}_\beta^{\;\alpha} + \bb{Q}_\alpha^{\;a}\,\bb{G}_a^{\;\alpha} - \bb{G}_a^{\;\alpha}\,\bb{Q}_\alpha^{\;a} \,, \qquad
\Delta(\bb{T}) = \bb{T}\otimes1+1\otimes\bb{T}+2\bb{T}^{\otimes2} \,, 
\end{align}
where $\bb{T}^{\otimes2}$ is the double-site version of $\bb{T}$. This operator commutes with all bosonic generators of the $\psuc$ algebra. In such a way the generators \eqref{Y(g)} are equivalent to
\begin{align}
\Delta(\wh{\bb{R}}_{a}^{\; b}) \, &= \wh{\bb{R}}_{a}^{\;b}\otimes1+1\otimes\wh{\bb{R}}_{a}^{\; b} + \tfrac{1}{2}\big[\bb{R}_{a}^{\;b}\otimes1,\bb{T}^{\otimes2}\big] \,, \el
\Delta(\wh{\bb{L}}_{\alpha}^{\enskip\beta}) &= \wh{\bb{L}}_{\alpha}^{\enskip\beta} \otimes 1+1 \otimes \wh{\bb{L}}_{\alpha}^{\enskip\beta} + \tfrac{1}{2}\big[\bb{L}_{\alpha}^{\enskip\beta}\otimes1,\bb{T}^{\otimes2}\big] \,, \label{Y(g)T}
\end{align}
However this elegant way of defining Yangian generators does not extend for supercharges $\wh{\bb{Q}}_\alpha^{\;a}$, $\wh{\bb{G}}^{\;\alpha}_{a}$ and central elements $\wh{\bb{C}}$, $\wh{\bb{C}}^\dg$, $\wh{\bb{H}}$. This is because the $\psuc$ algebra does not have a well-defined quadratic Casimir operator due to the degeneracy of the Killing--Cartan form. Finally, the evaluation representation is defined by \cite{BeisertYangian}
\begin{equation} \label{eq:ev_map_psuc}
{\rm ev}_u : \wh{\bb{J}}{}^a \mapsto i g u \, \bb{J}^a \,, 
\end{equation}
where $u$ is the rapidity \eqref{rapidity}.


\paragraph{Reflection.} 

In order to discuss the symmetry properties of the reflection matrices, we need to extend the reflected coproduct to Yangians. However, this is readily done by composing the reflection map $\kappa$ with the evaluation map ansatz \eqref{vJ_ansatz} giving
\begin{equation}
\kappa\bigl(\,\wh{\bb{J}}^a\bigr)\left|v\right\rangle =\gamma\,(-v+v_{0})\,\ul{\bb{J}}^a\left|v\right\rangle\,.
\end{equation}
Here $\kappa : v \mapsto -v$ and $\kappa : v_0, \gamma \mapsto v_0, \gamma$. The reflected coproduct of Yangian generators then straightforwardly follows by composing \eqref{eqn;coprodYangian} with $\kappa$. 


\subsection{\texorpdfstring{$Z=0$}{Z=0} giant graviton}\label{Sec:3.2}

The $Z=0$ giant graviton preserves all of the bulk Lie algebra. Thus the corresponding twisted Yangian is of the $\mbox{Y}(\g,\g)$ type and was presented in \cite{MR3}. Here we will give a more elegant form of this symmetry with the help of expressions \eqref{CasT} and \eqref{Y(g)T}.

Firstly, notice that in general
\begin{equation} \label{ComTJ}
[\bb{T},\wh{\bb{J}}^a] = \kappa_{bd} \,[\bb{J}^b \bb{J}^d,\wh{\bb{J}}^a] = f^{a}_{\; bc} \, (\wh{\bb{J}}^b {\bb{J}}^c + \bb{J}^c \wh{\bb{J}}^b) \,,
\end{equation}
where $\bb{T}$ is the Casimir operator. Recall that the $\psuc$ algebra does not have a well defined Casimir operator, as we have disused before, but has a well defined Casimir-like operator \eqref{CasT} which commutes with the bosonic generators. In such a way, bearing on the analogy to \eqref{Y(g)T}, we can combine the prescription \eqref{J2tw} with \eqref{ComTJ} giving level-2 twisted Yangian charges of the $Z=0$ giant graviton
\begin{align} \label{J2Z}
\dwt{\bb{R}}{}_{1}^{\; 2} := [\wh{\bb{R}}_{1}^{\;1},\wh{\bb{R}}_{1}^{\; 2}] + \tfrac{1}{4}\big[\wh{\bb{R}}{}_{1}^{\;2},\bb{T}\big] \,, \qquad
\dwt{\bb{L}}{}_{3}^{\;4} := [\wh{\bb{L}}_{3}^{\;3},\wh{\bb{L}}_{3}^{\;4}] + \tfrac{1}{4}\big[\wh{\bb{L}}{}_{3}^{\;4},\bb{T}\big] \,,
\end{align}
having coproducts given by
\begin{align} 
\Delta\big(\dwt{\bb{R}}{}_{1}^{\; 2}\big) &= \dwt{\bb{R}}{}_{1}^{\; 2}\otimes1+1\otimes\dwt{\bb{R}}{}_{1}^{\; 2} + [\wh{\bb{R}}{}_{1}^{\; 2}\otimes1,\bb{T}^{\otimes2}] \el 
& \quad + \tfrac{1}{4} \big[[{\bb{R}}{}_{1}^{\;1}\otimes1,\bb{T}^{\otimes2}],[{\bb{R}}{}_{1}^{\; 2}\otimes1,\bb{T}^{\otimes2}]\big] + \tfrac{1}{4}\big[[{\bb{R}}{}_{1}^{\; 2}\otimes1,\bb{T}^{\otimes2}],\bb{T}^{\otimes2}]\big] \,,  \label{J2Zcop1} \\
\Delta\big(\dwt{\bb{L}}{}_{3}^{\; 4}\big) &= \dwt{\bb{L}}{}_{3}^{\;4}\otimes1+1\otimes\dwt{\bb{L}}{}_{3}^{\;4} + [\wh{\bb{L}}{}_{3}^{\;4}\otimes1,\bb{T}^{\otimes2}] \el 
& \quad + \tfrac{1}{4} \big[[{\bb{L}}{}_{3}^{\;4}\otimes1,\bb{T}^{\otimes2}],[{\bb{L}}{}_{3}^{\;4}\otimes1,\bb{T}^{\otimes2}]\big] + \tfrac{1}{4}\big[[{\bb{L}}{}_{3}^{\;4}\otimes1,\bb{T}^{\otimes2}],\bb{T}^{\otimes2}]\big] \,,  \label{J2Zcop2}
\end{align}
where we have used \eqref{J2twcop} and \eqref{Y(g)T} implicitly. The rest of the Yangian algebra can be derived by commuting with the level-0 generators. 
The expressions given above have a relatively compact form, however the explicit form of the coproducts is very bulky. Also note that the terms with parameter $t$ are not present in \eqref{J2Z}. This is because the intertwining equation gives an additional constraint $t=0$. 

Finally, for finding the expressions of the reflected coproducts one has to use \eqref{cop_ref} together with
\begin{equation}
\Delta^\rfl(\wh{\bb{J}}^{A})=\ul{\wh{\bb{J}}}^{A}\otimes1+\mathcal{U}^{-[[A]]}
\otimes{\wh{\bb{J}}}^{A}+f_{\;
BC}^{A}\,\mathcal{U}^{-[[C]]}\,\ul{\bb{J}}^{B}\otimes{\bb{J}}^{C} .\label{
Y_refcoproduct}
\end{equation}
The boundary evaluation map is given by \cite{MR3}
\begin{equation} \label{eq:ev_map_B}
{\rm ev}_w : \wh{\bb{J}}{}^A \mapsto i g w \, \bb{J}^A \,, 
\end{equation}
where $w = \frac{i M}{2g}$ is the boundary spectral parameter. The same result may be obtained heuristically by applying the bulk-to-boundary map $x^\pm\mapsto \pm x_B$ to the bulk rapidity \eqref{rapidity} and using the boundary mass-shell condition \eqref{shortening_B},
\begin{equation}
u = x^+ + \frac{1}{x^+} - \frac{i M}{2g} \;\longmapsto\; x_B + \frac{1}{x_B} -
\frac{iM}{2g} = \frac{iM}{2g} = w \,.
\end{equation}
%


\paragraph{Symmetry constraints.}

By solving the reflection intertwining equation for all Lie algebra and Yangian symmetries, 
\begin{equation}
\Delta^{ref} (\bb{J}^A)\, K = K \,\Delta(\bb{J}^A) \,, \qquad \Delta^{ref}
(\,\dwt{\bb{J}}{}^A)\, K = K \,\Delta(\,\dwt{\bb{J}}{}^A) \,,
\end{equation}
one can obtain all reflection coefficients of any bound-state reflection matrix up to the overall dressing phase. This could be done in a similar way as it was done for the bound-state $S$-matrix in \cite{ALT}. However, due to very bulky form of the boundary Yangian, this would be extremely challenging.


\subsection{\texorpdfstring{$Z=0$}{Z=0} \texorpdfstring{$D7$}{D7}-brane: left factor}\label{sec:3.3}

The boundary Lie algebra for the left factor of the $Z=0$ $D7$-brane can be formally decomposed as $\hh = \g \backslash (\mm+\cc)$, where 
\begin{equation} \label{Xhm}
\hh = \{\bb{R}^{\;a}_{b},\, \bb{L}^{\;\alpha}_\beta,\;\bb{H}\}\,, \qquad 
\mm = \{\bb{Q}^{\;\alpha}_{b},\, \bb{G}^{\;a}_\beta\}\,, \qquad 
\cc = \{\bb{C},\;\bb{C}^\dg \}\,.
\end{equation}
This setup almost resembles the structure of a symmetric pair. In the latter case the boundary scattering would be governed by a twisted Yangian $\mbox{Y}(\g,\h)$ of type I \cite{DMS,VR} in a similar way as for the $Y=0$ giant graviton \cite{YangianY0}. Unfortunately, in the present case the symmetric pair structure breaks down due to the following relations,
\begin{equation} \label{CC}
\{ \bb{Q}_{\alpha}^{\; a},\bb{Q}_{\beta}^{\; b} \} =
\epsilon^{ab}\epsilon_{\alpha\beta}\,\bb{C}\,, \qquad\qquad
\{ \bb{G}_{a}^{\; \alpha},\bb{G}_{b}^{\; \beta} \} =
\epsilon^{\alpha\beta}\epsilon_{ab}\,\bb{C}^{\dg} \,.
\end{equation}
In other words, the presence of the central charges prevents us from applying the generic formalism discussed earlier. However, the algebra $\psuc$ has an $SL(2)$ outer automorphism, which is realized as a mixing of the supercharges. This automorphism can be used to rotate the central charges to a trivial point, $\bb{C}\equiv\bb{C}^\dag\equiv0$, in such a way the commutation relations \eqref{CC} in the rotated realization of the algebra are absent. We will use an analogue of this automorphism on the level of the twisted charges to construct the twisted Yangian.


\paragraph{Modified twisted Yangian $\Ygh$.} 
Let us first ignore the fact that the central charges $\bb{C}$ and $\bb{C}^\dagger$ are not symmetries of the boundary, and suppose they are in the boundary algebra $\hh$. Then following the prescription \eqref{twist}, and using the structural constants obtained from the Yangian Y$(\psuc)$, we obtain
\begin{align}
\wt{\bb{Q}}^{\prime\;a}_{\,\alpha} &= \wh{\bb{Q}}^{\;a}_{\alpha} + t_{_{Q}} \bb{Q}^{\;a}_{\alpha}  + \tfrac{1}{4}
\big(\bb{Q}_{\alpha}^{\; c}\,\bb{R}_{c}^{\; a} + \bb{R}_{c}^{\;
a}\,\bb{Q}_{\alpha}^{\; c} + \bb{Q}_{\gamma}^{\; a} \,\bb{L}_{\alpha}^{\;
\gamma} + \bb{L}_{\alpha}^{\; \gamma}\,\bb{Q}_{\gamma}^{\; a} +
\bb{H}\,\bb{Q}_{\alpha}^{\; a} -
2\,\varepsilon_{\alpha\gamma}\varepsilon^{ac}\,\bb{C}\,\bb{G}_{c}^{\;
\gamma}\big) \el
& = \wh{\bb{Q}}^{\;\alpha}_{b} + t_{_{Q}} {\bb{Q}}^{\;\alpha}_{b} + \tfrac{1}{4}
\big(\bb{H}\,\bb{Q}_{\alpha}^{\; a} - [\bb{T}^{\hh},{\bb{Q}}^{\;\alpha}_{b}] -
2\,\varepsilon_{\alpha\gamma}\varepsilon^{ac}\,\bb{C}\,\bb{G}_{c}^{\;
\gamma}\big) \,, \el
\wt{\bb{G}}_{\,a}^{\prime\; \alpha} &= \wh{\bb{G}}_{a}^{\; \alpha} - t_{_{G}} \bb{G}_{a}^{\; \alpha} -
\tfrac{1}{4} \big( \bb{G}_{c}^{\; \alpha}\,\bb{R}_{a}^{\; c} + \bb{R}_{a}^{\; c}\,
\bb{G}_{c}^{\; \alpha} + \bb{G}_{a}^{\; \gamma}\,\bb{L}_{\gamma}^{\; \alpha} +
\bb{L}_{\gamma}^{\; \alpha}\, \bb{G}_{a}^{\; \gamma} + \bb{H}\,\bb{G}_{a}^{\;
\alpha} - 2\,\varepsilon_{ac}\varepsilon^{\alpha\gamma}\,\bb{C}^{\dg}
\bb{Q}_{\gamma}^{\; c} \big) \el
& = \wh{\bb{G}}_{a}^{\; \alpha} - t_{_{G}} \bb{G}_{a}^{\; \alpha} -
\tfrac{1}{4} \big( \bb{H}\,\bb{G}_{a}^{\;\alpha} - [\bb{T}^{\hh},{\bb{Q}}^{\;\alpha}_{b}] - 2\,\varepsilon_{ac}\varepsilon^{\alpha\gamma}\,\bb{C}^{\dg}
\bb{Q}_{\gamma}^{\; c} \big) \,, \label{QGt}
\end{align} 
where $\bb{T}^{\hh}$ is the Casimir-like operator \eqref{CasT} restricted to the subalgebra $\hh$ \eqref{Xhm}. The coproducts of these twisted charges are
\begin{align}
\Delta(\wt{\bb{Q}}^{\prime\;a}_{\,\alpha}) &= \wt{\bb{Q}}^{\prime\;a}_{\,\alpha}
\otimes 1 + \mathcal{U}^{+1}\!\otimes \wt{\bb{Q}}^{\prime\;a}_{\,\alpha} +
\bb{Q}_{\alpha}^{\; c}\otimes\bb{R}_{c}^{\; a} + \bb{Q}_{\gamma}^{\;
a}\otimes\bb{L}_{\alpha}^{\; \gamma} + \tfrac{1}{2}\bb{Q}_{\alpha}^{\;
a}\otimes\bb{H} - \varepsilon_{\alpha\gamma}\varepsilon^{ac}\,\bb{G}_{c}^{\;
\gamma}\,\mathcal{U}^{+2}\!\otimes\bb{C} \,, \el
\Delta(\wt{\bb{G}}_{\,a}^{\prime\; \alpha}) &= \wt{\bb{G}}_{\,a}^{\prime\;
\alpha}\otimes1 + \mathcal{U}^{-1}\!\otimes\wt{\bb{G}}_{\,a}^{\prime\; \alpha} -
\bb{G}_{c}^{\; \alpha}\otimes\bb{R}_{a}^{\; c} - \bb{G}_{a}^{\;
\gamma}\otimes\bb{L}_{\gamma}^{\; \alpha} - \tfrac{1}{2}\bb{G}_{a}^{\;
\alpha}\otimes\bb{H} +
\varepsilon_{ac}\varepsilon^{\alpha\gamma}\,\bb{Q}_{\gamma}^{\;
c}\,\mathcal{U}^{-2}\!\otimes\bb{C}^{\dg} .
\end{align} 
As expected, we see that these charges violate the coideal property due to central charges acting on the boundary. We can overcome this problem by adding a twist resembling the $SL(2)$ automorphism, 
\begin{align}
\wt{\bb{Q}}^{\;a}_{\alpha} &= \wt{\bb{Q}}^{\prime\;a}_{\,\alpha} 
+
\varepsilon_{\alpha\gamma}\varepsilon^{ac}\,(\bb{C}-g \alpha)\,\bb{G}_{c}^{\;
\gamma} \,, \el
\wt{\bb{G}}_{a}^{\; \alpha} &= \wt{\bb{G}}_{\,a}^{\prime\; \alpha} 
-
\varepsilon_{ac}\varepsilon^{\alpha\gamma}\,(\bb{C}^{\dg}-g \alpha^{-1})
\bb{Q}_{\gamma}^{\; c}\,. \label{QGt2}
\end{align} 
The coproducts of the new charges are then readily found to be
\begin{align}
\Delta(\wt{\bb{Q}}^{\;a}_{\alpha}) &= \wt{\bb{Q}}^{\;\alpha}_{b} \otimes 1 + \mathcal{U}^{+1}\otimes
\wt{\bb{Q}}^{\;\alpha}_{b} + \bb{Q}_{\alpha}^{\; c}\otimes\bb{R}_{c}^{\; a} +
\bb{Q}_{\gamma}^{\; a}\otimes\bb{L}_{\alpha}^{\; \gamma} +
\tfrac{1}{2}\,\bb{Q}_{\alpha}^{\; a}\otimes\bb{H} \,, \el 
\Delta(\wt{\bb{G}}_{a}^{\; \alpha}) &= \wt{\bb{G}}_{a}^{\; \alpha}\otimes1 +
\mathcal{U}^{-1}\!\otimes\wt{\bb{G}}_{a}^{\; \alpha} - \bb{G}_{c}^{\;
\alpha}\otimes\bb{R}_{a}^{\; c} - \bb{G}_{a}^{\;
\gamma}\otimes\bb{L}_{\gamma}^{\; \alpha} - \tfrac{1}{2}\,\bb{G}_{a}^{\;
\alpha}\otimes\bb{H}\,, \label{DG_final}
\end{align}
and thus the coideal property \eqref{coideal_g_h} is satisfied. 

The parameters $t_{_{Q}}$ and $t_{_{G}}$ in the twist \eqref{QGt} are constrained by requiring the twisted central charges
\begin{equation}
\bb{\dwt{C}} = \epsilon_{ab}\epsilon^{\alpha\beta}\{ \wt{\bb{Q}}^{\;a}_{\alpha} ,
\, \wt{\bb{Q}}^{\;b}_{\beta} \}  \,, \qquad\qquad 
\bb{\dwt{C}}{}^\dagger =
\epsilon_{\alpha\beta}\epsilon^{ab}\{ \wt{\bb{G}}^{\;\alpha}_{a} , \,
\wt{\bb{G}}^{\;\beta}_{b} \} \,,
\end{equation}
to be coreflective. This gives a constraint $t_{_{Q}}\!=t_{_{G}}\!=\sqrt{g^2+1/4}$. The square root may be eliminated by using the fundamental mass-shell condition $x_B + 1/x_B = i/g$ \eqref{shortening_B}. In such a way we obtain the very elegant expression, $t_{_{Q}}\!=t_{_{G}}\!= i g/x_B + 1/2$. 


\paragraph{Symmetry constraints.}

The complete reflection matrix $K$ \eqref{KXref} then follows from simple symmetry arguments. Indeed,
\begin{align}
\big( K \,\wt{\Q}^{\; 1}_3 - \ul{\wt{\Q}}{}^{\; 1}_3 \, K \big)\,| k \rangle^1 = 0
\qquad{\rm and}\qquad \big( K \, \wt{\Q}^{\; 1}_3 - \ul{\wt{\Q}}{}^{\; 1}_3 \, K \big)\,| k
\rangle^2 = 0 
\end{align}
lead to the reflection coefficients that coincide with \eqref{XBC} as required.


\section{Quantum affine boundary algebras}\label{sec:4}

In this section we will consider a $q$-deformed model of the boundary scattering from the $Z=0$ giant graviton and the left factor of the $Z=0$ $D7$-brane considered earlier. We will start by briefly reclling the construction of the quantum affine coideal subalgebras \cite{VR} (see \cite{Le} for explicit details on the non-affine coideal subalgebras) and the bound-state representation of the quantum affine algebra $\afQ$ \cite{BGM,LMR}. We will then construct the corresponding boundary algebras using the same approach as for the $q$-deformed model of the reflection from the $Y=0$ giant graviton \cite{LMR2}. 


\subsection{Quantum affine coideal subalgebras}
 
Let the quantum deformed universal enveloping algebra $\Uq$ of a semisimple Lie algebra $\g$ of rank $n$ be generated by the elements $E_i,\, F_i,\, K_{i}^{\pm1}$ ($K_i=q^{H_i}$, $i=1,\ldots,n$), that correspond to the standard Drinfeld-Jimbo realization. 
The Hopf algebra structure of $\Uq$ is given by 
\begin{align}
\label{cop1}
& \Delta (K_i) = K_i \otimes K_i \,,  && S(K^{-1}_i) = K_i\,, && \epsilon (K_i) = 1\,, \el
& \Delta (E_i) = E_i \otimes 1 + K^{-1}_i \otimes E_i\,,  && S(E_i) = - K_i E_i\,, && \epsilon(E_i) =0 \,, \el
& \Delta (F_i) = F_i \otimes K_i + 1 \otimes F_i\,, && S(F_i) = - F_i K^{-1}_i\,, && \epsilon(F_i) =0 \,.
\end{align}
Being a Hopf algebra, $\Uq$ admits a right adjoint actions that makes $\Uq$ into a right module. The right adjoint action is given by 
\begin{align}
\left({\rm ad}_r \,E_i \right) \!A &= (-1)^{[A][E_i]} K_i A E_i - K_i E_i A \,, \el
\left({\rm ad}_r \,F_i \right) \!A &= (-1)^{[A][F_i]}A F_i - F_i K^{-1}_i \!A K_i \,, 
& \left({\rm ad}_r \,K^{-1}_i \right) \!A &= K_i A K^{-1}_i \,,
\end{align}
where $(-1)^{[A][E_i]}$ and $(-1)^{[A][F_i]}$ are the fermionic grade factors.
We shall also be using a short--hand notation $\big({\rm ad}_r \,E_i \cdots E_j \big) A = \big({\rm ad}_r \,E_i \cdots {\rm ad}_r \,E_j \big) A$ and simi\-larly for $F_i$. 

Let $\Ua$ be the universal enveloping algebra of $\hat\g$, the affine extension of $\g$.
Let $\pi=\{\alpha_{1},\alpha_{2},\ldots,\alpha_{n}\}$ be the set of simple positive roots of $\g$, and let $\wh\pi = \alpha_0\cup\pi$, where $\alpha_0$ denotes the affine root. Let $E_0,\, F_0,\,K_0^{\pm1}$ be the affine generators of $\Ua$, and let $\mathcal{T}$ denote the abelian subgroup $\mathcal{T}\subset\Ua$ generated by all $K_i^{\pm1}$ and $K_0^{\pm1}$.

Consider an involution $\theta$ of $\hat\g$ such that the associated root space automorphism $\Theta$ may be represented by
\begin{equation} \label{AutDef1}
\Theta (\alpha_0) \in -\alpha_{p(0)} - \bb{Z} (\pi\backslash\alpha_{p(0)}) \quad\text{and}\quad \Theta (\alpha_i) = \alpha_i \, \quad \text{for all} \quad \alpha_i \in \pi_\Theta=\pi\backslash\alpha_{p(0)}\,.
\end{equation}
where $p(0)\in\{0,1,\ldots,n\}$, and satisfying
\begin{equation} \label{AutDef2}
\alpha_0-\Theta(\alpha_0) = k \delta\,, \qquad\text{where}\qquad \begin{cases}
k=1 \text{ for } p(0)\neq0 \,,\\ k=2 \text{ for } p(0)=0 \,,\end{cases}
\end{equation}
where $\delta$ is the imaginary root;recall that $\alpha_0 = \delta - \vartheta$, where $\vartheta$ is the highest root. Then $\Theta$ induces a subalgebra $\mathcal{M}\subset\Ua$ generated by $E_i$, $F_i$ and $K_i^\pm$ for all $\alpha_i\in\pi_\Theta$ and a $\Theta$--fixed subgroup $\mathcal{T}_\Theta$. 
Furthermore, there exists a sequence $\{\alpha_{i_{1}},\ldots,\alpha_{i_{r}}\}$, $\alpha_{i_k}\in\pi_\Theta$, and a set of positive integers $\{m_{1},\ldots,m_{r}\}$ such that the algebra elements defined by
\begin{align} \label{twistB}
\wt{E}_0 &= F_0 K_0^{-1} - d_{y} \,\tth(F_{0}) K_0^{-1} \,, \qquad \tth(F_{0}) = \bigl(\mbox{ad}_{r}\,{E_{i_{1}}}^{(m_{1})}\!\cdots {E_{i_{r}}}^{(m_{r})}\bigr)E'_{p(0)} \,,\el
\wt{F}_0 &= E'_0 K_0^{-1} - d_{x} \,\tth(E'_{0}) K_0^{-1} \,, \qquad \tth(E'_{0}) = \bigl(\mbox{ad}_{r}\,{F_{i_{1}}}^{(m_{1})}\!\cdots {F_{i_{r}}}^{(m_{r})}\bigr)F_{p(0)} \,,
\end{align}
where $E'_i = E_i K_i$, together with $\mathcal{T}_\Theta$, $\mathcal{M}$ and suitable $d_{x},\,d_{y}\in\bb{C}$ generate a quantum affine coideal subalgebra $\afB\subset\Ua$ which is compatible with the reflection equation. Note that quite often the boundary algebra includes all of the Cartan subgroup $\mathcal{T}$. In such cases the factor of $K_0^{-1}$ in \eqref{twistB} can be omitted. The boundary algebras we will be considering in the next sections will be exactly of this type. The case with $p(0)=0$ will correspond to the $q$-deformed model of the $Z=0$ giant graviton, while the $p(0)\neq0$ case will correspond to the left factor of the $D7$-brane. 
\paragraph{Example.} 
Here we will give a simple example illustrating the technique described above. Consider $\mathcal{U}_q(\hat{\mathfrak{sl}}_2)=\{E_i, F_i, K_i^{\pm1} |\, i=0,1\}$, a quantum affine extension of the Lie algebra $\g=\mathfrak{sl}_2$. The set of simple roots in this case is $\hat\pi=\{\alpha_0, \alpha_1\}$, where $\alpha_1$ is the regular root and $\alpha_0$ is the affine root. There are two boundary scattering problems associated with this algebra that are relevant to us. 

{\it 1.\ Case with $\mathit{p(0)=0}$.} Consider a boundary which respects all of the bulk Lie algebra, i.e.\ the boundary Lie algebra is $\hh=\g=\mathfrak{sl}_2$. This means that the root $\alpha_1$ is respected by the boundary, while $\alpha_0(=\delta-\alpha_1)$ is not (by definition). Then the associated root space automorphism by \eqref{AutDef1} and \eqref{AutDef2} is constrained to
\begin{equation}
\Theta_1 (\alpha_1) = \alpha_1, \qquad \alpha_0 - \Theta_1 (\alpha_0) = 2 \delta \qquad \text{giving} \qquad\Theta_1 (\alpha_0) = - \alpha_0 - 2 \alpha_1 \,.
\end{equation}
Hence the non-affine subalgebra of the boundary algebra $\hat{\mathcal{B}}$ is $\mathcal{M}=\{E_1, F_1\}$, while the affine part, by \eqref{twistB}, is generated by the twisted affine generators 
\begin{align}
\wt{E}_0 &= F_0 - d_{y} \,\tth(F_{0}) \,, \qquad \tth(F_{0}) = \bigl(\mbox{ad}_{r}\,E_{1}^{2}\bigr)E'_{0} \,,\el
\wt{F}_0 &= E'_0 - d_{x} \,\tth(E'_{0}) \,, \qquad \tth(E'_{0}) = \bigl(\mbox{ad}_{r}\,F_{1}^{2}\bigr)F_{0} \,,
\end{align}
for suitable $d_x, d_y$\,. 

{\it 2.\ Case with $\mathit{p(0)\neq0}$.} Consider a boundary which respects none of the bulk Lie algebra, i.e.\ the boundary Lie algebra consists of the Cartan subalgebra only. This means that both roots, $\alpha_1$ and $\alpha_0$, are not respected by the boundary. Then
\begin{equation}
\alpha_0 - \Theta_2 (\alpha_0) = \delta \qquad \text{giving} \qquad\Theta_2 (\alpha_0) = - \alpha_1 \,.
\end{equation}
Hence the boundary algebra $\hat{\mathcal{B}}$ is generated by the Cartan subalgebra and twisted affine generators 
\begin{align}
\wt{E}_0 &= F_0 - d_{y} \,\tth(F_{0}) \,, \qquad \tth(F_{0}) = E'_{1} \,,\el
\wt{F}_0 &= E'_0 - d_{x} \,\tth(E'_{0}) \,, \qquad \tth(E'_{0}) = F_{1} \,,
\end{align}
with suitable $d_x, d_y$\,. For more details on these coideal subalgebras see \cite{VR}.


\subsection{Quantum affine algebra of the \texorpdfstring{$q$}{q}-deformed worldsheet scattering}


The symmetry algebra $\widehat{\mathcal{Q}}$ of the $q$-deformed worldsheet scattering is a deformation of the centrally extended affine algebra $\widehat{\mathfrak{sl}}(2|2)$ \cite{BGM}. It is generated by four sets of the Chevalley generators $E_i$, $F_i$, $K_i$ ($i=1,\,2,\,3,\,4$) and two sets of central elements $U_{k},V_{k}$ ($k=2,\,4$) with $U_{k}$ being responsible for the braiding of the coproduct.  The set of simple positive roots is $\wh{\pi}=\pi \cup \alpha_4=\{\alpha_1,\,\alpha_2,\,\alpha_3,\,\alpha_4\}$, where $\alpha_4$ is the affine root. The roots $\alpha_1$ and $\alpha_3$ are bosonic, while $\alpha_2$ and $\alpha_4$ are fermionic.

Let us start by recalling the symmetric matrix $DA$ and the normalization matrix $D$ associated to the Cartan matrix $A$ for $\widehat{\mathfrak{sl}}(2|2)$:
\begin{equation}
DA=\begin{pmatrix}2 & -1 & 0 & -1\\
-1 & 0 & 1 & 0\\
0 & 1 & -2 & 1\\
-1 & 0 & 1 & 0
\end{pmatrix},\qquad D=\mathrm{diag}(1,-1,-1,-1).\label{DA}
\end{equation}
The algebra is then defined accordingly by the following non-trivial commutation relations,
\begin{align}
 & K_{i}E_{j}=q^{DA_{ij}}E_{j}K_{i}, &  & K_{i}F_{j}=q^{-DA_{ij}}F_{j}K_{i},\el
 & \{E_{2},F_{4}\}=-\tilde{g}\tilde{\alpha}^{-1}(K_{4}-U_{2}U_{4}^{-1}K_{2}^{-1}), &  & \{E_{4},F_{2}\}=\tilde{g}\tilde{\alpha}^{+1}(K_{2}-U_{4}U_{2}^{-1}K_{4}^{-1}),\el
 & [E_{j},F_{j}\}=D_{jj}\frac{K_{j}-K_{j}^{-1}}{q-q^{-1}}, &  & [E_{i},F_{j}\}=0,\quad i\neq j,\ i+j\neq6. \label{Qcomrel}
\end{align}
These are supplemented by a set of Serre relations ($j=1,\,3$) 
\begin{align}
 & [E_{j},[E_{j},E_{k}]]-(q-2+q^{-1})E_{j}E_{k}E_{j}=0, && [E_{1},E_{3}]=E_{2}E_{2}=E_{4}E_{4}=\{E_{2},E_{4}\}=0, \el
 & [F_{j},[F_{j},F_{k}]]-(q-2+q^{-1})F_{j}F_{k}F_{j}=0, && [F_{1},F_{3}]=F_{2}F_{2}=F_{4}F_{4}=\{F_{2},F_{4}\}=0.
\end{align}
and the central elements are related to the quartic Serre relations (for $k=2,\,4$) as follows 
\begin{align}
 & \{[E_{1},E_{k}],[E_{3},E_{k}]\}-(q-2+q^{-1})E_{k}E_{1}E_{3}E_{k}=g\alpha_{k}(1-V_{k}^{2}U_{k}^{2}),\el
 & \{[F_{1},F_{k}],[F_{3},F_{k}]\}-(q-2+q^{-1})F_{k}F_{1}F_{3}F_{k}=g\alpha_{k}^{-1}(V_{k}^{-2}-U_{k}^{-2})\,.
\end{align}
This algebra has three central charges,
\begin{align}
&C_{1} = K_{1}K_{2}^{2}K_{3} \,, && C_{2} = g\alpha_{2}(1-V_{2}^{2}U_{2}^{2}) \, ,&& C_{3} = g\alpha_{2}^{-1}(V_{2}^{-2}-U_{2}^{-2}) \,, \label{C123}
\end{align}
plus three affine counterparts of theirs. Finally, the central elements $V_{k}$ are constrained by the relation $K_{1}^{-1}K_{k}^{-2}K_{3}^{-1}=V_{k}^{2}.$ 


\paragraph{Hopf algebra.}

The group-like elements $X\in\{1,K_{j},U_{k},V_{k}\}$ ($j=1,2,3,4$ and $k=2,4$) have the coproduct $\Delta$ defined in the usual way, $\Delta(X)=X\otimes X$, while for the remaining Chevalley generators they are deformed by the central elements $U_{k}$
\begin{equation}
\Delta(E_{j})=E_{j}\otimes1+K_{j}^{-1}U_{2}^{+\delta_{j,2}}U_{4}^{+\delta_{j,4}}\otimes E_{j}\,,\quad\Delta(F_{j})=F_{j}\otimes K_{j}+U_{2}^{-\delta_{j,2}}U_{4}^{-\delta_{j,4}}\otimes F_{j}\,.\label{copEF}
\end{equation}


\paragraph{Representation.}

We shall be using the $q$-oscillator representation (for any complex $q$ not a root of unity) constructed in \cite{LMR}. The bound-state representation is defined on vectors 
\begin{equation}
|m,n,k,l\rangle=(\mathsf{a}_{3}^{\dag})^{m}(\mathsf{a}_{4}^{\dag})^{n}(\mathsf{a}_{1}^{\dag})^{k}(\mathsf{a}_{2}^{\dag})^{l}\,|0\rangle\,,
\end{equation}
where the indices $1,\,2$ denote bosonic and $3,\,4$ - fermionic oscillators; the total number of excitations $k+l+m+n=M$ is the bound-state number and the dimension of the representation is dim$\,=\!4M$.  This representation constrains the central elements as $U:=U_2=U^{-1}_4$  and $V:=V_2=V^{-1}_4$ and describes an excitation with momentum $p$ defined by the relation $U^2={\rm e}^{ip}$. 

The triples corresponding to the bosonic and fermionic $\mathfrak{sl}_{q}(2)$ in this representation are given by 
\begin{align}
 & H_{1}|m,n,k,l\rangle=(l-k)\,|m,n,k,l\rangle\,, &  & H_{3}|m,n,k,l\rangle=(n-m)\,|m,n,k,l\rangle\,,\el
 & E_{1}|m,n,k,l\rangle=[k]_{q}\,|m,n,k-1,l+1\rangle\,, &  & E_{3}|m,n,k,l\rangle=|m+1,n-1,k,l\rangle\,,\el
 & F_{1}|m,n,k,l\rangle=[l]_{q}\,|m,n,k+1,l-1\rangle\,, &  & F_{3}|m,n,k,l\rangle=|m-1,n+1,k,l\rangle\,.
\end{align}
The supercharges act on basis states as 
\begin{align}
H_{2}|m,n,k,l\rangle= & ~-\left\{ C-\frac{k-l+m-n}{2}\right\} |m,n,k,l\rangle\,,\el
E_{2}|m,n,k,l\rangle= & ~a~(-1)^{m}[l]_{q}\,|m,n+1,k,l-1\rangle+b~|m-1,n,k+1,l\rangle\,,\el
F_{2}|m,n,k,l\rangle= & ~c~[k]_{q}\,|m+1,n,k-1,l\rangle+d~(-1)^{m}\,|m,n-1,k,l+1\rangle\,.
\end{align}
Here $[n]_q=(q^n-q^{-n})/(q-q^{-1})$ denotes the $q$-number and $C$ is the $q$-factor of the central element $V = q^C$ and represents the energy of the state. The representation labels $a,b,c,d$ satisfy constraints 
\begin{align}
 & ad=\frac{q^{\frac{M}{2}}V-q^{-\frac{M}{2}}V^{-1}}{q^{M}-q^{-M}}\,, && bc=\frac{q^{-\frac{M}{2}}V-q^{\frac{M}{2}}V^{-1}}{q^{M}-q^{-M}}\,, \el
 & ab=\frac{g\alpha}{[M]_{q}}(1-U^{2}V^{2})\,, && cd=\frac{g\alpha^{-1}}{[M]_{q}}(V^{-2}-U^{-2})\,, \label{rep}
\end{align}
which altogether give the multiplet shortening (mass-shell) condition 
\begin{equation} \label{short}
\frac{g^{2}}{[M]_{q}^{2}}(V^{-2}-U^{-2})(1-U^{2}V^{2})=\frac{(V-q^{M}V^{-1})(V-q^{-M}V^{-1})}{(q^{M}-q^{-M})^{2}}\;.
\end{equation}
The explicit $x^{\pm}$ parametrization of the representation labels is 
\begin{align}
a & =\sqrt{\frac{g}{[M]_{q}}}\gamma\,, && b=\sqrt{\frac{g}{[M]_{q}}}\frac{\alpha}{\gamma}\frac{x^{-}\!-x^{+}}{x^{-}}\,,\el
c & =\sqrt{\frac{g}{[M]_{q}}}\frac{\gamma}{\alpha\, V}\frac{i\,\tilde{g}\, q^{\frac{M}{2}}}{g(x^{+}\!+\xi)}\,, &  & d=\sqrt{\frac{g}{[M]_{q}}}\frac{\tilde{g}\, q^{\frac{M}{2}}V}{i\, g\,\gamma}\frac{x^{+}\!-x^{-}}{\xi x^{+}\!+1}\,.\label{abcd_q}
\end{align}
The central elements in this parametrization read as
\begin{align}
&U^2 = \frac{1}{q^M} \frac{x^+ + \xi}{x^- + \xi} =  q^M \frac{x^+}{x^-}\frac{\xi x^- + 1}{\xi x^+ + 1}, 
  && V^2 = \frac{1}{q^M} \frac{\xi x^+ + 1}{\xi x^- + 1} = q^M \frac{x^+}{x^-}\frac{x^- + \xi}{x^+ + \xi}, \label{UVrep}
\end{align}
while the shortening condition \eqref{short} becomes
\begin{equation}
 \frac{1}{q^{M}}\left(x^+ + \frac{1}{x^+}\right)-q^M\left(x^- + \frac{1}{x^-}\right) = \left(q^M-\frac{1}{q^M}\right) \left(\xi+\frac{1}{\xi}\right),
\end{equation}
where $\xi = -i \tilde g (q-q^{-1})$ and $\tilde g^2={g^2}/({1-g^2(q-q^{-1})^2})$.

The action of the affine charges $H_{4}$, $E_{4}$, $F_{4}$ is defined in exactly the same way as for the regular supercharges subject to the following substitutions, $C\to-C$ and \mbox{$(a,b,c,d)\to(\tilde{a},\tilde{b},\tilde{c},\tilde{d})$}.  The affine labels $\tilde{a},\tilde{b},\tilde{c},\tilde{d}$ are acquired from (\ref{abcd}) by a simple replacement 
\begin{equation}
V\rightarrow V^{-1}\,, \qquad 
x^{\pm}\rightarrow\frac{1}{x^{\pm}}\,, \qquad 
\gamma\rightarrow\frac{i\tilde{\alpha}\gamma}{x^{+}}\,, \qquad 
\alpha\rightarrow\alpha\,\tilde{\alpha}^{2}\,. 
\end{equation}
The multiplicative spectral (evaluation) parameter of the algebra is
\begin{equation}\label{z}
z = \frac{1-U^2V^2}{V^2-U^2} \,.
\end{equation}
%


\paragraph{Reflection.}
Recall that reflection maps incoming states $|m,n,k,l\rangle\in V(p)$ to outgoing (reflected) states $|m,n,k,l\rangle\in V(-p)$ while keeping the boundary invariant \eqref{eq:reflection}
\begin{equation}
K : V(p) \otimes V(s) \longrightarrow V(-p) \otimes V(s) \,. \nonumber
\end{equation}
The representation defined in the paragraph above describes incoming states with momentum $p$ given by the relation ${\rm e}^{ip}=U^2$. Then the representation corresponding to the reflected states with momentum $-p$ will have the central element equal to ${\rm e}^{-ip}=U^{-2}$, i.e. reflection acts by inverting the central element $U \mapsto U^{-1}$. The conservation of the total number of fermions and bosons together with the energy conservation constrains the central element $V$ and Cartan generators $K_i$ to be invariant under the reflection.
This implies that there is a reflection automorphism $\kappa : \afQ \to \afQ^{ref}$ of the algebra defined by
\begin{equation}
\kappa:(V,U)\mapsto(\underline{V},\underline{U}) \qquad\mbox{and}\qquad \kappa:(E_{j},F_{j},K_{j})\mapsto(\underline{E}_{j},\underline{F}_{j},\underline{K}_{j}) \,,
\end{equation}
where the underlined elements generate the reflected algebra $\afQ^{ref}$. Then the constraints
\begin{equation} \label{refUVK}
\ul{U}=U^{-1} \,, \qquad \ul{V}=V \,, \qquad \ul{K}_i=K_i \,,
\end{equation}
define the representation of the reflected algebra. The representation labels $\underline{a},\,\underline{b}\,,\underline{c},\,\underline{d}$
associated to the charges $\underline{E}_{j},\underline{F}_{j}$ are obtained by replacing $U\mapsto U^{-1}$ in \eqref{rep} and similarly for the affine ones. Hence the labels of
the reflected charges are related to the initial ones as
\begin{equation}
\underline{a}=\frac{\,\ul{\gamma}\,}{\gamma}a,\qquad\underline{b}=\frac{\gamma\alpha^{2}}{\ul{\gamma}}\frac{cd}{a}V^{2},\qquad\underline{c}=\frac{\ul{\gamma}}{\gamma\alpha^{2}}\frac{ab}{d}V^{-2},\qquad\underline{d}=\frac{\,\gamma\,}{\ul{\gamma}}d,\label{ref_abcd}
\end{equation}
giving 
\begin{align}
\underline{a} &=\sqrt{\frac{g}{[M]_{q}}}\ul{\gamma}, && \underline{b}=\sqrt{\frac{g}{[M]_{q}}}\frac{\alpha}{\ul{\gamma}}\frac{\tilde{g}^{2}(x^{+}\!-x^{-})}{g^{2}(1+\xi x^{-})(\xi+x^{+})}, \el
\underline{c} & =\sqrt{\frac{g}{[M]_{q}}}\frac{\ul{\gamma}}{\alpha\, V}\frac{gq^{\frac{M}{2}}(\xi x^{-}\!+1)}{i\tilde{g}\, x^{-}}, && \underline{d}=\sqrt{\frac{g}{[M]_{q}}}\frac{\tilde{g}\, q^{\frac{M}{2}}V}{i\, g\,\ul{\gamma}}\frac{x^{+}\!-x^{-}}{\xi x^{+}\!+1}, \label{abcd_ref}
\end{align}
The extension to the affine case is straightforward. Here we have chosen
$\underline{a}=\frac{\,\ul{\gamma}\,}{\gamma}a$ as an initial
constraint with $\ul{\gamma}$ being the reflected version
of $\gamma$, i.e.\ $\kappa(\gamma)=\ul{\gamma}$. The reflection map for the 
$x^\pm$ parametrization is found by comparing 
\eqref{abcd_ref} with \eqref{abcd_q}, giving
\begin{equation}
\kappa:x^{\pm}\mapsto-\frac{x^{\mp}+\xi}{\xi x^{\mp}+1}\,.
\end{equation}
It is involutive, $\kappa^2=id$, and is in agreement with the one conjectured in \cite{MN}%
\footnote{The authors of \cite{MN} are using the $x^{\pm}$ parametrization
of \cite{BK}, while we use the one of \cite{BGM}. The map between
these two is $x_{\text{\tiny{BK}}}^{\pm}=g{\tilde{g}}^{-1}(x_{\text{\tiny{BGM}}}^{\pm}+\xi)$.%
}.
In the $q\to1$ limit this maps specializes to the usual reflection map, $\kappa:x^{\pm}\mapsto-x^{\mp}$, as required.

Let us also introduce the reflected coproducts of $E_i$ and $F_i$ , 
\begin{equation}
\Delta^{\! ref\!}(E_{j})=\ul{E}_{j}\otimes1+K_{j}^{-1}U^{-\delta_{j,2}+\delta_{j,4}}\otimes {E}_{j}\,,
 \quad \Delta^{\! ref\!}(F_{j})=\ul{F}_{j}\otimes K_{j}+U^{+\delta_{j,2}-\delta_{j,4}}\otimes {F}_{j}\,,
\end{equation}
where as in the previous section, $\Delta^{ref} := (\kappa \otimes 1) \circ \Delta$ and we have used \eqref{refUVK} implicitly.
These shall play an important role in finding the explicit form of the reflection matrix.

The expressions in \eqref{ref_abcd} may be casted in a matrix form
\begin{align}
\begin{pmatrix}\underline{a}&\underline{b}\\ \underline{c}&\underline{d} \end{pmatrix} D
=T\begin{pmatrix}a&b\\ c&d \end{pmatrix}T^{-1} 
\quad\text{with}\quad
D=\begin{pmatrix} \gamma/\ul{\gamma} & 0 \\ 0 & \ul{\gamma}/\gamma \end{pmatrix}, 
\quad
T=\begin{pmatrix} U^{-2} & 0 \\ 0 & -z \end{pmatrix}, \label{abcd_qref}
\end{align}
revealing the explicit relation between two isomorphic representations of $\afQ$. Here were treat $\gamma$ and $\ul{\gamma}$ as unconstrained parameters defining the representations of incoming and reflected states. In the $q\to1$ limit \eqref{abcd_qref} specializes to \eqref{abcd1_ref} as required.


\subsection{\texorpdfstring{$q$}{q}-deformed \texorpdfstring{$Z=0$}{Z=0} giant graviton}

The $Z=0$ giant graviton preserves all of the bulk Lie algebra. Therefore the corresponding $q$-deformed model of this boundary preserves all regular charges and all Cartan subalgebra $\mathcal{T}$ of $\afQ$. The affine generators $E_4$ and $F_4$ are not preserved by the boundary itself, but give rise to the twisted affine generators of the quantum affine coideal subalgebra $\afB_Z\subset\afQ$ .  


\paragraph{Coideal subalgebra.}

The boundary conditions define the root space automorphism $\Theta_Z$ associated to this boundary to act on the simple roots as 
\begin{equation} \label{autoZ}
\Theta_Z(\alpha_{i}) =\alpha_{i} \quad\text{for}\quad i = 1,2,3, \qquad\text{and}\qquad
\Theta_Z(\alpha_{4}) =-\alpha_4-2\alpha_3-2\alpha_2-2\alpha_1 \,.
\end{equation}
Thus $\pi_{\Theta_Z}=\{\alpha_{1},\,\alpha_{2},\,\alpha_{3}\}$
and it gives rise to the subalgebra $\mathcal{M}_Z$ of $\widehat{\mathcal{Q}}$ .
The affine part of the boundary algebra $\afB_Z$ is generated by the twisted affine charges
\begin{align}
 & \tE_{312}=F_{4} - d_{y}\,\tth(F_{4}) \,, & & \tth(F_{4})=\left(\mbox{ad}_{r} E_1 E_3 E_2 E_3 E_2 E_1 \right)E_{4}' \,,\label{twE321Z}\\
 & \tF_{312}=E_{4}' - d_{x}\,\tth(E_{4}') \,, & & \tth(E_{4}')=\left(\mbox{ad}_{r} F_1 F_3 F_2 F_3 F_2 F_1 \right)F_{4} \,,\label{twF321Z}
\end{align}
where the action of $\tth$ is induced by \eqref{autoZ}.
Any other non-trivial ordering of the generators in the adjoint action above is equivalent up to a sign. Here by non-trivial we assume the obtained operator is non-zero.
Note that the form of the twisted charges above slightly differs from those in \eqref{twistB} because the boundary respects all of the Cartan subalgebra $\mathcal{T}\subset\afQ$.
The rest of $\afB_Z$ can be furnished with the help of the right adjoint action of $\mbox{ad}_{r}\mathcal{M}_Z$,
\begin{align}
\wt{E}_{12} &= \left(\mbox{ad}_{r}F_{3}\right)\wt{E}_{312}\,, & \wt{F}_{12} & =\left(\mbox{ad}_{r}E_{3}\right)\wt{F}_{312}\,, \\
\wt{E}_{32} &= \left(\mbox{ad}_{r}F_{1}\right)\wt{E}_{312}\,, & \wt{F}_{32} & =\left(\mbox{ad}_{r}E_{1}\right)\wt{F}_{312}\,, \\
\wt{E}_{2} &= \left(\mbox{ad}_{r}F_{1}F_{3}\right)\wt{E}_{312}\,, & \wt{F}_{2} & =\left(\mbox{ad}_{r}E_{1}E_{3}\right)\wt{F}_{312}\,, \label{twEF2Z}\\
\wt{C}_{2} &= \left(\mbox{ad}_{r}E_{2}\right)\wt{E}_{312}\,, & \wt{C}_{3} & =\left(\mbox{ad}_{r}F_2\right)\wt{F}_{312}\,. \label{twC23Z}
\end{align}

Let us show the coideal property for the these charges explicitly. It is enough to show the coideal property for a pair of twisted affine charges only. For simplicity reasons we choose \eqref{twEF2Z},

\begin{align} \label{eq:copZ1}
\Delta(\wt{E}_{2}) &= (\mbox{ad}_{r}\, F_{1}F_{3})F_{4} K_4^{-1}\otimes K_{13}-d_y(\mbox{ad}_{r}\, E_{2}E_{3}E_{2}E_{1})E_{4}'K_4^{-1}\otimes K_{2321} + U K_4^{-1}\otimes\wt{E}_{2} \el
& \quad + (q-q^{-1})\Big(q^{-1}F_{4}K_4^{-1}\otimes K_{4}\left[F_{1}, F_{3}\right]_{q^{2}} \el
& \qquad -q(\mbox{ad}_{r}\, F_{1})F_{4}\otimes K_{14}F_{3}+q^{-1}(\mbox{ad}_{r}\, F_{3})F_{4}\otimes K_{34}F_{1}\Big) \el
& \quad -d_y(q-q^{-1})(U\tp1) \Big(q^{-2}U E_{4}'\otimes K_{4}\,\big\{ E_{2}',\big[E_{3}',\left[E_{1}', E_{2}'\right]_{q}\big]_{q^{3}}\big\} \el
& \qquad -U(\mbox{ad}_{r}\, E_{1})E_{4}'\otimes K_{14}(\mbox{ad}_{r}\, E_{2}E_{3})E_{2}'-U(\mbox{ad}_{r}\, E_{3})E_{4}'\otimes K_{34}(\mbox{ad}_{r}\, E_{2}E_{1})E_{2}' \el
& \qquad +(\mbox{ad}_{r}\, E_{2}E_{3})E_{4}'\otimes K_{234}(\mbox{ad}_{r}\, E_{2})E_{1}'+(\mbox{ad}_{r}\, E_{2}E_{1})E_{4}'\otimes K_{214}(\mbox{ad}_{r}\, E_{2})E_{3}' \el
& \qquad +(\mbox{ad}_{r}\, E_{2}E_{1}E_{3})E_{4}'\otimes K_{2134}E_{2}'\Big) \el
& \in \afQ\otimes\afB_Z\,,
\end{align}
and similarly for $\Delta(\wt{F}_{2})$.
%
%
The short-hand notation $K_{i\ldots j} = K_i \cdots K_j $ has been employed, and $[A,B]_{q^n} = AB-q^n BA$ is the $q$-deformed commutator. The coideal property for the rest of the twisted affine charges follows from the ${\rm ad}_r \mathcal{M}_Z$-invariance of $\afB_Z$. 


\paragraph{Boundary representation.} 

The next step is to construct the boundary bound-state representation of the coideal subalgebra $\afB_Z$. The constraints defining the representation are the commutation relations in the third line of \eqref{Qcomrel}, and the coreflectivity of the regular central charges $C_2$, $C_3$ \eqref{C123} and the twisted affine central charges $\wt{C}_2$, $\wt{C}_3$ \eqref{twC23Z}. We will start by constructing the boundary representation of the regular supercharges $E_2$ and $F_2$ and the central element $V$. We will denote the latter as $V_B$ in order to distinguish it from the bulk representation \eqref{UVrep}. Note that the deformation parameter $U$ is {\it not} in the boundary algebra and thus does not have a boundary representation. This can be easily seen by inspecting \eqref{eq:copZ1}, 
 $U$ never appears in the right factor of the tensor product. In such a way the algebra constraints \eqref{C123} get modified for the boundary algebra. 

The algebra constraints for $C_2$ and $C_3$ for incoming and reflected states in the bulk are given by
\begin{align}
C_{2} \otimes 1 & = g\alpha(1-U^{2}V^{2}) \otimes 1\,, & C_{3} \otimes 1 & = g\alpha^{-1}(V^{-2}-U^{-2}) \otimes 1\,, \el
\ul{C}_{2} \otimes 1 & = g\alpha(1-U^{-2}V^{2}) \otimes 1\,, & \ul{C}_{3} \otimes 1 & = g\alpha^{-1}(V^{-2}-U^{2}) \otimes 1\,.
\end{align}
Here we have used \eqref{refUVK} implicitly and the tensor space structure is $\,bulk\otimes boundary$. 
Then requiring their coproducts
\begin{align}
\Delta (C_{2}) &= C_{2}\otimes1+V^{2}U^{2}\otimes C_{2}\,, & \Delta(C_{3}) &= C_{3}\otimes V_B^{-2}+U^{-2}\otimes C_{3}\,,\el
\Delta^{\rfl}(C_{2}) &= \ul{C}_{2}\otimes1+V^{2}U^{-2}\otimes {C}_{2}\,, & \Delta^\rfl(C_{3}) &= \ul{C}_{3}\otimes V_B^{-2}+U^{2}\otimes {C}_{3}\,,
\end{align}
to be coreflective, $\Delta(C_{i})=\Delta^\rfl(C_{i})$, we find the boundary algebra constraints for the regular central charges to be
\begin{equation}
1\otimes C_{2} = 1\otimes {g\alpha}\,, \qquad\qquad 1\otimes C_{3} = 1\otimes{g\alpha^{-1}}V_B^{-2}\,.
\end{equation}
Therefore the representation constraints for the boundary algebra are
\begin{align}
a_{B}d_{B} & =\frac{q^{\frac{M}{2}}V_{B}-q^{-\frac{M}{2}}V_{B}^{-1}}{q^{M}-q^{-M}}\,, 
& b_{B}c_{B} & =\frac{q^{-\frac{M}{2}}V_{B}-q^{\frac{M}{2}}V_{B}^{-1}}{q^{M}-q^{-M}}\,,\el
a_{B}b_{B} & = \frac{g\alpha}{[M]_q} \,, 
& c_{B}d_{B} & = \frac{g\alpha^{-1}}{[M]_q} V_{B}^{-2}\,.
\end{align}
These relations force the boundary labels to be
\begin{align}
a_{B} &= \sqrt{\tfrac{g}{[M]_q}}\gamma_{B}\,, 
& b_{B} &= \sqrt{\tfrac{g}{[M]_q}}\frac{\alpha}{\gamma_{B}},\el
c_{B} &= \sqrt{\tfrac{g}{[M]_q}}\frac{\gamma_{B}}{\alpha}\frac{i\tilde{g}}{g\xi}\frac{q^{M/2}\left(1-q^{-M}V_{B}^{2}\right)}{V_{B}}\,, 
& d_{B} &= \sqrt{\tfrac{g}{[M]_q}}\frac{\tilde{g}}{\gamma_{B}}\frac{q^{M/2}\left(V_{B}^{2}-q^{-M}\right)}{ig\xi\, V_{B}}\,, \label{Bconstraints}
\end{align}
where $V_B$ is required to satisfy
\begin{equation}
\left(V_{B}^{2}-q^{-M}\right)\left(V_{B}^{2}-q^{M}\right)=\frac{\xi^{2}}{\xi^{2}-1}\,.\label{VB2}
\end{equation}
A convenient parametrization satisfying this constraint is
\begin{equation} \label{VB}
V_{B}^{2}=q^{M}\frac{x_{B}}{x_{B}+\xi}=q^{-M}\frac{1+\xi x_{B}}{1-\xi^{2}}\,.
\end{equation}
In this way the boundary labels become
\begin{align}
a_{B} &= \sqrt{\tfrac{g}{[M]_q}}\gamma_{B} \,, 
& b_{B} &= \sqrt{\tfrac{g}{[M]_q}}\frac{\alpha}{\gamma_{B}}\,,\el
c_{B} &= \sqrt{\tfrac{g}{[M]_q}}\frac{\gamma_{B}}{\alpha}\frac{i\tilde{g}}{g}\frac{q^{M/2}}{V_{B}(x_{B}+\xi)}\,, 
& d_{B} &= \sqrt{\tfrac{g}{[M]_q}}\frac{\tilde{g}}{i g\gamma_{B}}\frac{V_B q^{M/2}\left(x_{B}+\xi\right)}{\xi x_B + 1}\,. \label{abcd_Bq}
\end{align}
Consequently, the mass-shell constraint 
\begin{equation}
\left(a_{B}d_{B}-q^{M}b_{B}c_{B}\right)\left(a_{B}d_{B}-q^{-M}b_{B}c_{B}\right)=1\,,
\end{equation}
in this parametrization becomes
\begin{equation}
\frac{q^{-2 M} g^2 \left(1+x_B^2+2 x_B \xi \right)^2}{[M]_q^2 \, (\xi ^2-1) \, x_B^2}=1\,.
\end{equation}
In the $q\to1$ limit it gives the usual (non-deformed) mass-shell constraint
\begin{equation}
-\frac{g^{2}}{M^{2}}\left(x_{B}+\frac{1}{x_{B}}\right)^{2}=1\qquad\Longrightarrow\qquad x_{B}+\frac{1}{x_{B}}=\frac{iM}{g}\,.
\end{equation}
Furthermore, the $q\to1$ limit gives $V_B\to1$, and labels \eqref{abcd_Bq} reproduce the usual non-deformed boundary labels \eqref{abcd_B}, as required. 

Let us now turn to the construction of the boundary representation labels of the affine generators $E_4$ and $F_4$. We will construct the affine representation in a similar way as we did for the regular one above, except we will not give the explicit details of the coreflectivity of the twisted affine central charges as we did for the regular ones. This is because the explicit form of the coproducts of $\wt{C}_2$ and $\wt{C}_3$ is very large and thus we will only state the final constraints we have obtained.

The representation constraints that follow from the commutation relations \eqref{Qcomrel} give
\begin{equation}
\tl{a}_{B}\tl{d}_{B} = \frac{q^{\frac{M}{2}}\wt{V}_{B}-q^{-\frac{M}{2}}\wt{V}_{B}^{-1}}{q^{M}-q^{-M}}\,, \qquad\qquad
\tl{b}_{B}\tl{c}_{B} = \frac{q^{-\frac{M}{2}}\wt{V}_{B}-q^{\frac{M}{2}}\wt{V}_{B}^{-1}}{q^{M}-q^{-M}}\,. 
\end{equation}
Bearing on the analogy to the affine bulk labels we choose the following ansatz for the affine boundary labels,
\begin{align}
\tl{a}_{B} &= \sqrt{\tfrac{g}{[M]_q}}\frac{\gamma_{B}\tl{\alpha}}{A_B} \,, 
& \tl{b}_{B} &= \sqrt{\tfrac{g}{[M]_q}}\frac{\alpha\tl{\alpha}}{\gamma_{B}}B_B\,,\el
\tl{c}_{B} &= \frac{q^{-\frac{M}{2}}\wt{V}_{B}-q^{\frac{M}{2}}\wt{V}_{B}^{-1}}{(q^{M}-q^{-M})\,\tl{b}_B}\,,
& \tl{d}_{B} &= \frac{q^{\frac{M}{2}}\wt{V}_{B}-q^{-\frac{M}{2}}\wt{V}_{B}^{-1}}{(q^{M}-q^{-M})\,\tl{a}_B}\,, \label{abcd_ABq_ansatz}
\end{align}
where $A_B$ and $B_B$ are undetermined parameters. Then using this ansatz and requiring $\wt{C}_2$ and $\wt{C}_3$ to be coreflective we find additional constraints that solve this requirement,
\begin{equation} \label{ABq_sol}
A_B = -i\, x_B \,, \qquad B_B = - i (x_B + 2 \xi) \,, \qquad {V}_B^2 \, \wt{V}_B^2 =1 +  \frac{\xi^2}{\xi^2-1}\,.
\end{equation}
These define the affine boundary labels to be
\begin{align}
\tl{a}_{B} &= \sqrt{\tfrac{g}{[M]_q}}\frac{i \gamma_{B}\tl{\alpha}}{x_B} \,, 
& \tl{b}_{B} &= \sqrt{\tfrac{g}{[M]_q}}\frac{\alpha\tl{\alpha}}{i\gamma_{B}}(x_B+2\xi)\,,\el
\tl{c}_{B} &= -\sqrt{\tfrac{g}{[M]_q}} \frac{\tl{g}\,q^{\frac{M}{2}}\gamma_B}{g \alpha \tl{\alpha} (1+\xi x_B) \wt{V}_B} \,,
& \tl{d}_{B} &= \sqrt{\tfrac{g}{[M]_q}} \frac{\tl{g}\,q^{-\frac{M}{2}}}{g\tl{\alpha} \gamma_B \wt{V}_B}\frac{1-\xi(x_B+2\xi)}{\xi^2-1} \,. \label{abcd_ABq}
\end{align}
The coreflectivity property also constrains the parameters $d_y$ and $d_x$ to be
\begin{equation} 
d_y = (\alpha \tl{\alpha})^{-2}\,, \qquad \qquad d_x = -(\alpha \tl{\alpha})^2 \,,
\end{equation}
thus fixing the last undetermined elements of $\afB_Z$.
%


Finally we want to give two useful relations of the boundary representation that are closely linked to those of the bulk representation. Namely, the evaluation parameter $z$ may be expressed in terms of the bulk representation labels as
\begin{equation}
z=\frac{g}{\tl g\,\alpha\,\tl\alpha}(a\tl b-b\tl a)\,, \qquad\qquad
z^{-1}=\frac{g\,\alpha\,\tl\alpha}{\tl g}(c\tl d-d\tl c)\,.
\end{equation}
In a similar way, for the boundary representation, we obtain
\begin{align}
q^M = \frac{g}{\tl g\,\alpha\,\tl\alpha}(a_B\tl b_B-b_B\tl a_B)\,, \qquad
q^{-M} = 
  V_B \wt{V}_B\frac{g\,\alpha\,\tl\alpha}{\tl g}(c_B\tl d_B-d_B\tl c_B)\,.
\end{align}
In the $q\to1$ limit parameter $z$ can be expanded in series as $z=1-2ighu+\mathcal{O}(h^2)$, where $q\sim e^h$ and $u$ is given by \eqref{rapidity}. The second term in this expansion reveals the Yangian evaluation map of \eqref{eq:ev_map_psuc}. Similarly for the boundary case we obtain $q^M = 1 - 2 i g h w + \mathcal{O}(h^2)$, where $w=iM/g$, and is in a perfect agreement with the boundary evaluation map \eqref{eq:ev_map_B} as required.


\paragraph{Symmetry constraints.}

The boundary algebra $\afB_Z$ allows us to find any bound-state reflection matrix up to the overall dressing phase by solving the boundary intertwining equation \eqref{eq:int_B}
\begin{equation} 
\Delta^{ref} (J^A)\, K_q = K_q \,\Delta(J^A) \qquad\text{for all}\qquad J^A\in\afB_Z \,,\nonumber
\end{equation}
This can be done in a similar way as in \cite{LMR}, where the bound-state $S$-matrix for the algebra $\afQ$ was found. However these calculations are rather complicated and thus we will reduce our goal to finding the analytic expressions of the reflection matrices with the total bound-state number $M\leq3$. These are the fundamental reflection matrix $K_q^{Aa}$ and the bound-state reflection matrices $K_q^{Ba}$ and $K_q^{Ab}$. Here indices $^A$ and $^B$ denote the fundamental and $M=2$ bound-states in the bulk, the indices $^a$ and $^b$ in the same way denote the boundary (bound-) states. These matrices in the explicit form are given in the Appendix \ref{AppA}. We have checked that they are unitary and satisfy the reflection equation. Also we have calculated some higher order bound-state reflection matrices numerically, and checked that they satisfy the reflection equation.


\subsection{\texorpdfstring{$q$}{q}-deformed \texorpdfstring{$Z=0$}{Z=0} \texorpdfstring{$D7$}{D7}-brane: left factor}

The left factor of the $Z=0$ $D7$-brane does not respect any of the Lie supercharges $\Q_a^{\;\alpha}$, $\G_\alpha^{\;a}$ or central charges $\C$, $\C^\dagger$ \eqref{Xhm}. Hence the corresponding $q$-deformed model of this boundary in addition to the affine supercharges $E_4$ and $F_4$ does not respect the regular supercharges $F_2$ and $E_2$ (and central elements $C_2$, $C_3$). These generators combined together will give rise to the twisted affine generators of the quantum affine coideal subalgebra $\afB_X \subset \afQ$ . The boundary is a singlet with respect to the boundary algebra, thus we will not need to construct the boundary representation of $\afB_X$.


\paragraph{Coideal subalgebra.}

The boundary conditions define the root space automorphism $\Theta_X$ associated to the left factor of the $D7$-brane to act on the simple roots as 
\begin{align}
\Theta_X(\alpha_{1}) &=\alpha_{1}, & \Theta_X(\alpha_{2})&=-\alpha_4-\alpha_1-\alpha_3,\el
\Theta_X(\alpha_{3}) &=\alpha_{3}, & \Theta_X(\alpha_{4})&=-\alpha_2-\alpha_1-\alpha_3.
\end{align}
Thus $\pi_{\Theta_X}=\{\alpha_{1},\,\alpha_{3}\}$
and it gives rise to the subalgebra $\mathcal{M}_X$ of $\widehat{\mathcal{Q}}$ .

As in the previous case, we build $\afB_X$ based on the affine extension,
hence $p(4)=2$. This setup fixes the twisted affine charges to be 
\begin{align}
 & \tE_{312}=F_{4} - d_{y}\,\wt{\theta}(F_{4})\,, & & \wt{\theta}(F_{4})=\left(\mbox{ad}_{r}E_{3}E_{1}\right)E_{2}'\,,\label{twE321}\\
 & \tF_{312}=E_{4}' - d_{x}\,\wt{\theta}(E_{4}')\,, & & \wt{\theta}(E_{4}')=\left(\mbox{ad}_{r}F_{3}F_{1}\right)F_{2}\,.\label{twF321}
\end{align}
Let us show the coideal property for the these charges explicitly, 
\begin{align}
\Delta(\wt{E}_{312}) &= F_{4}\otimes K_{4}-d_{y}(\mbox{ad}_{r}\, E_{3}E_{1})E_{2}'\otimes K_{312}-U\otimes\widetilde{E}_{312}\el
& \quad -d_{y}(q-q^{\!-1})\Big((\mbox{ad}_{r}\, E_{1})E_{2}'\otimes K_{12}E_{3}' \el
& \qquad\qquad\qquad\qquad -(\mbox{ad}_{r}\, E_{3})E_{2}'\otimes K_{32}E_{1}'+q^{\!-1}E_{2}'\otimes K_{2}[E_{1}',E_{3}']_{q^{2}}\!\Big)\quad\el
 & \in \afQ\otimes\afB_X,
\end{align}
and
\begin{align}
\Delta(\wt{F}_{312}) &= E_{4}\otimes K_{4}-d_{x}(\mbox{ad}_{r}\, F_{3}F_{1})F_{2}\otimes K_{312}-U^{-1}\otimes\widetilde{F}_{312}\el
& \quad -d_{x}\,q^{\!-1}(q-q^{\!-1})\Big((\mbox{ad}_{r}\, F_{3})F_{2}\otimes K_{32}F_{1} \el
& \qquad\qquad\qquad\qquad\qquad -q^2(\mbox{ad}_{r}\, F_{1})F_{2}\otimes K_{12}F_{3} + F_{2}\otimes K_{2}[F_{1},F_{3}]_{q^{2}}\!\Big)\el
 & \in \afQ\otimes\afB_X.
\end{align}
The rest of $\afB_X$ can be furnished with the help of the right adjoint action of $\mbox{ad}_{r}\mathcal{M}_X$,
\begin{align}
\wt{E}_{12} & =\left(\mbox{ad}_{r}F_{3}\right)\widetilde{E}_{312}, & \wt{F}_{12} & =\left(\mbox{ad}_{r}E_{3}\right)\widetilde{F}_{312}, \\
\wt{E}_{32} & =\left(\mbox{ad}_{r}F_{1}\right)\widetilde{E}_{312}, & \wt{F}_{32} & =\left(\mbox{ad}_{r}E_{1}\right)\widetilde{F}_{312}, \\
\wt{E}_{2} & =\left(\mbox{ad}_{r}F_{1}F_{3}\right)\wt{E}_{312}, & \wt{F}_{2} & =\left(\mbox{ad}_{r}E_{1}E_{3}\right)\wt{F}_{312}. \label{twEF1}
\end{align}
As previously, the coideal property for these charges is obvious since $\afB_X$ is invariant under the adjoint action of $\mathcal{M}_X$. 

The final ingredients of $\afB_X$ are the twisted affine central charges $\dwt{C}_2$ and $\dwt{C}_3$ that can be obtained by anticommuting two twisted affine charges, e.g.
\begin{align}
\dwt{C}_2 & = \{ \wt{E}_{12}\,, \wt{E}_{32} \}\,, \qquad\qquad \dwt{C}_3 = \{ \wt{F}_{12}\,, \wt{F}_{32} \}. \label{twC23}
\end{align}
These twisted affine central charges must be reflective. And because the boundary is a singlet we require $\dwt{\ul{C}}_{2}=\dwt{C}_{2}$ and $\dwt{\ul{C}}_{3}=\dwt{C}_{3}$. This gives us the following constraints,
\begin{equation}
1+d_x \chi (q+q^{-1}) - \frac{d_x^2 \chi^2}{\xi^2-1}  = 0 \,, \qquad
\frac{1}{\xi^2-1} + \frac{d_y}{\chi} (q+q^{-1}) - \frac{d_y^2}{\chi^{2}} =0 \,,
\end{equation}
where $\chi = \dfrac{\tl{g}}{g \alpha\tl{\alpha}}\,$. These constraints can be solved by introducing the following ansatz,
\begin{equation}
d_{y}= \frac{\tilde{g}}{g \, \alpha \tilde{\alpha}} \, V'_B \qquad\mbox{and}\qquad d_{x}= -\frac{g\,\alpha\tilde{\alpha}}{\tilde{g}} \,V'_B (1-\xi^2)\,,
\end{equation}
where
\begin{equation}
V'_B = q\frac{1-\xi x'_B}{1-\xi^2} = q^{-1}\frac{x'_B}{x'_B-\xi} \,.
\end{equation}
Note that $V'_B$ is related to $V_B$ in \eqref{VB} by setting $M=1$ and inverting the deformation parameter, $q\rightarrow q^{-1}$, giving $\xi \rightarrow -\xi$. Thus $x'_B$ may be understood as the spectral parameter of the {\it oppositely} deformed fundamental boundary.%
\footnote{It is possible to choose a parametrization of $d_x$ and $d_y$ that it
would agree with the one used for the $Z=0$ giant graviton, i.e.\ in terms of
$x_B$, not $x'_B$. However this would make expressions of the reflection
matrices much more complicated and the pole structure would not be transparent.}


\paragraph{Symmetry constraints.}
The structure of the $q$-deformed reflection matrix is equivalent to the non-deformed case \eqref{KX} and the corresponding vector space is the same \eqref{KXbasis}. The bosonic charges $E_{1}$, $F_{1}$ and $E_{3}$, $F_{3}$ constrain the reflection matrix to be diagonal,
\begin{align}
K_q\,|k\rangle^{1} & = A_q \,|k\rangle^{1},
& K_q\,|k\rangle^{2} & = B_q \,|k\rangle^{2},
& K_q\,|k\rangle^{\alpha} & = C_q \,|k\rangle^{\alpha},
\end{align}
and we have added the subscript $_q$ to distinguish the $q$-deformed reflection matrix from the one in \eqref{KXref}. Next we choose the normalization for the reflection of the state $|k\rangle^{1}$ to be $A_q = 1$. Then the intertwining equation for the charge $\tE_{2}$ gives
\begin{equation}
\big(\K_q\, \tE_{2}-\ul{\tE}_{2}\,K_q\big)|k\rangle^{1}=0
\qquad\Longrightarrow\qquad B_q = \frac{x'_B + x^+}{x'_B + \kappa(x^+)}\frac{\,\ul{\gamma}\,}{\gamma}.
\end{equation}
Equivalently, the same constraint may be found by considering the reflection of states $|0\rangle^{\alpha}$ and employing the charge $\tF_{2}$. Next we consider the reflection of the $|k\rangle^{2}$ state. The intertwining equation in this case leads to
\begin{equation}
\big(K_q\, \tE_{2}-\ul{\tE}_{2}\,K_q\big)|k\rangle^{2}=0
\quad\Longrightarrow\quad C_q = \frac{(1\!+\!\xi x^-)(1\!+\!\xi x^+)}{1-\xi^2}\frac{(1 + x'_B \kappa(x^-))(x'_B + x^+)}{(1 + x'_B x^-)(x'_B + \kappa(x^+))}\frac{\,\ul{\gamma}^2}{\gamma^2}.
\end{equation}
%

Let us perform some consistency checks. It is straightforward to check that this
reflection matrix satisfies the unitarity condition $K_q(p)K_q(-p)=1$. In the
$q\to1$ limit the $q$-deformed reflection coefficients $A_q$, $B_q$ and $C_q$
specialize to the non-deformed ones given in \eqref{XBC} as required. Finally
we also explicitly verified that it satisfies the reflection equation when the total bound
state number $M\leq5$. Thus it is good indication to expect it to hold for any $M$.


\section{Discussion}\label{sec:5}

In this work we have constructed the twisted Yangian describing the boundary symmetries and the worldsheet boundary scattering of the left factor of the open string attached to the $Z=0$ $D7$-brane in the $\ads$ background. This was the last unknown boundary symmetry algebra and now all of the (generalized) twisted Yangians for the well known AdS/CFT boundaries have been constructed. We have also given an elegant form of the Yangian generators of the $Z=0$ giant graviton.

We have computed the $q$-deformed analogues of the reflection matrices corresponding to the aforementioned $D7$-brane and the $Z=0$ giant graviton. The latter was earlier considered in \cite{MN}. In this language we have found a rather compact way of expressing the corresponding symmetry algebras as the coideal subalgebras of the quantum affine algebra $\afQ$ constructed in \cite{BGM}. 

We have explicitly calculated the $q$-deformed fundamental ($K_q^{Aa}$) and two-particle bound-state ($K_q^{Ba}$ and $K_q^{Ab}$) reflection matrices of the $Z=0$ giant graviton. We have checked that these reflection matrices obey both non-affine and twisted-affine symmetries and satisfy the reflection equation. We have also performed these tests for some higher order bound-state reflection matrices which we have calculated numerically only. The analytic form of the generic bound-state reflection matrix could be found using the same approach as in \cite{LMR}, where the bound-state $S$-matrix for the $q$-deformed worldsheet scattering was found. The matrix structure of the $S$-matrix and the reflection matrix of the $Z=0$ giant graviton is of the same form, however the boundary algebra is of a much more complicated structure than the bulk one, thus finding the generic bound-state reflection matrix would be a highly complicated exercise and goes beyond of the scope of the present work. 

The coideal subalgebras we have constructed in the rational $q\to1$ limit are required to reproduce the corresponding twisted Yangian algebras. We have checked that the subalgebra $\afB_X$ of the $q$-deformed left factor of the $Z=0$ $D7$-brane reproduces its twisted Yangian. However finding the rational limit of the subalgebra $\afB_Z$ of the $q$-deformed $Z=0$ giant graviton is rather involved and leads to a complicated combination of the level-2 and level-0 Yangian generators. Thus we have limited our goal to checking if the $q$-deformed reflection matrices in the $q\to1$ limit reproduce the regular ones and we found this to be in a perfect agreement. 

The $q$-deformed boundaries we have considered support only some subalgebra $\afB\subset\afQ$. One could ask if it would be possible to construct such an integrable boundary that it would support {\it all} of the algebra $\afQ$. Our answer is that this is {\it not possible}. This is because the commutation relations of $\afQ$ and the coreflectivity of the (non-twisted) central charges, both regular and affine, can not be satisfied simultaneously. Interestingly, for the representations of the $\psuc$ algebra there is a simple bulk-to-boundary map. However for the $q$-deformed case we do not see any obvious bulk-to-boundary map.  

Some interesting further questions that would be worthwhile to investigate would include the role of the secret symmetry and finding the $q$-deformed analogue of the achiral boundary \cite{MR4}. This would include the important question of constructing the diagonal embedding of the quantum deformed algebras. Also certain boundary scattering matrices were shown to display extra symmetries \cite{RegelskisSecret} corresponding to the so-called secret symmetry of the $S$-matrix \cite{MMT}. The secret symmetry is also a symmetry of the of the $q$-deformed $S$-matrix \cite{deLeeuwSecret} and it would be interesting to see if the $q$-deformed $K$-matrices accommodate the corresponding extra symmetries. 


\paragraph{Acknowledgements.}

The authors would like to thank Niklas Beisert, Takuya Matsumoto and Alessandro Torrielli for useful discussions. We are grateful to Niall MacKay for many comments and suggestions on the manuscript. M.\ dL.\ thanks Swiss National Science Foundation for funding under project number 200021-137616. V.R.\ also thanks the UK EPSRC for funding under grant EP/H000054/1.\\



\appendix

\section{\texorpdfstring{$q$}{q}-deformed reflection matrices}\label{AppA}

In this Appendix we present the explicit forms of the $q$-deformed reflection matrices for the $Z=0$ giant graviton. We enumerate the basis for fundamental particles as
\begin{align}
&e_1 = |0,0,1,0\rangle\,, &&e_2 = |0,0,0,1\rangle , &&e_3 = |1,0,0,0\rangle , &&e_4 = |0,1,0,0\rangle\,.
\end{align}
and two-particle bound states as
\begin{align}
&\hat{e}_1  = |0,0,2,0\rangle\,, &&\hat{e}_2  = |0,0,1,1\rangle\,, &&\hat{e}_3  = |0,0,0,2\rangle &&\hat{e}_4  = |1,0,1,0\rangle\el
&\hat{e}_5  = |1,0,0,1\rangle\,, &&\hat{e}_6  = |0,1,1,0\rangle\,, &&\hat{e}_7  = |0,1,0,1\rangle &&\hat{e}_8  = |1,1,0,0\rangle\,.
\end{align}
We will use the symbol ``$\,\circ\,$'' to denote the tensor product of states to keep the expressions as compact as possible. For the same reason we will also omit writing the subindex $q$. Our normalization is such that $K \, e_1\tc e_1 = e_1\tc e_1$ and equivalently for the bound-states. We have checked that these reflection matrices satisfy the reflection equation and unitarity requirement, $K(-p)K(p)=1$. 


\paragraph{Reflection matrix $K^{Aa}_q$}

\begin{align}
 K \, e_a \tc e_a &= \, e_a \tc e_a \,,\el
 K \, e_a \tc e_\alpha &= k_3 \, e_\alpha \tc e_a+k_2 \, e_a \tc e_\alpha \,,\el
 K \, e_\alpha \tc e_a &= k_8 \, e_\alpha \tc e_a+k_4 \, e_a \tc e_\alpha \,,\el
 K \, e_\alpha \tc e_\alpha &= k_9 \, e_\alpha \tc e_\alpha \,,\el
 K \, e_1 \tc e_2 &= k_1 \, e_2 \tc e_1+(1-q^{-1}{k_1}) \, e_1 \tc e_2-q^{-1}k_6 \, e_4 \tc e_3+q^{-2}k_6 \, e_3 \tc e_4 \,,\el
 K \, e_2 \tc e_1 &= (1-q k_1) \, e_2 \tc e_1+k_1 \, e_1 \tc e_2+k_6 \, e_4 \tc e_3-q^{-1}k_6 \, e_3 \tc e_4 \,,\el
 K \, e_3 \tc e_4 &= -q k_5 \, e_2 \tc e_1+k_5 \, e_1 \tc e_2+k_7 \, e_4 \tc e_3+(-q^{-1}k_7+k_9) \, e_3 \tc e_4 \,,\el
 K \, e_4 \tc e_3 &= q^2 k_5 \, e_2 \tc e_1-q k_5 \, e_1 \tc e_2+(-q k_7+k_9) \, e_4 \tc e_3+k_7 \, e_3 \tc e_4 \,.\qquad\qquad\qquad\quad
\end{align}
Here $a=1,2$ and $\alpha=3,4$ , and the coefficients $k_i$ are
\begin{align}
&k_1=  \left[\frac{ U^2 (\xi +x^+)-q (\xi +x^-)}{x_B-x^-} - 
U^2 (1- U^2V^2) \frac{x_B+\xi}{x_B-x^-} \frac{x^--\kappa(x^-)}{\xi +x^+ }\right]\! \frac{V^2}{U^2} \,,\qquad\qquad\quad\;\el
&k_2 = \frac{q (\xi + x_B)-U^2 (\xi +x^+)}{q U^2 (x_B-x^-)} \,,\el
&k_3 = q^{\frac{1}{2}} (1-U^2 V^2) \frac{x^--\kappa(x^-)}{x^--x_B}\frac{V}{U} \frac{{\gamma_B}}{\gamma}\,,\el
&k_4 = zq^{-\frac{1}{2}} \frac{(x^--\kappa(x^-))(x_B+\xi)}{(x_B-x^-)(\xi +x^-)}\frac{V}{U} \frac{\,\ul{\gamma}\,}{\gamma_B}\,,\el
&k_5 =\frac{q^{-\frac{3}{2}}}{\alpha} \left[
\frac{q \left(\xi +x^-\right)-U^2 \left(\xi +x^+\right)}{\left(x_B-x^-\right)}+
\frac{z \left(x^+-\kappa \left(x^+\right)\right) \left(x_B+\xi \right)}{q^2 \left(x_B-x^-\right) \left(\xi +x^-\right)}\right] \frac{V}{U} \,\ul{\gamma}\gamma_B \,,\el
&k_6 = \alpha q^{\frac{1}{2}} \frac{U^4-1}{U^2} \left[
q V^2 \frac{\xi +x^+}{x_B-x^-}  + \frac{(1-U^2 V^2)(x_B+\xi)}{x^--x_B} \right]\frac{V}{U}\frac{1}{\gamma \gamma_B} \,,\el
&k_7 =\left[ \frac{z V^2}{q} \frac{U^4-1}{U^2} \frac{ x_B + \xi }{x_B-x^-} + \frac{(1-U^2 V^2) (x^+-\kappa(x^+))}{x_B-x^-}\right] \frac{\,\ul{\gamma}\,}{\gamma} \,,\el
&k_8 = \frac{z U^2  (x_B+\xi ) + \xi +x^-}{x^--x_B} \frac{\,\ul{\gamma}\,}{\gamma}\,,\el
&k_9 = z\frac{x_B-\kappa(x^-)}{x^--x_B} \frac{\,\ul{\gamma}\,}{\gamma} \,.
\end{align}
%


\paragraph{Reflection matrix $K^{Ba}_q$}

\begin{align}
 K \, \hat{e}_1 \tc e_1 &= \, \hat{e}_1 \tc e_1 \,,\el
 K \, \hat{e}_1 \tc e_2 &= q k_3 \, \hat{e}_2 \tc e_1-\tfrac{k_5}{q} \, \hat{e}_8 \tc e_1 + \big(1-k_3-\tfrac{k_3}{q^2}\big) \, \hat{e}_1 \tc e_2-\tfrac{k_4}{q} \, \hat{e}_6 \tc e_3 + \tfrac{k_4}{q^2} \, \hat{e}_4 \tc e_4 \,,\el
 K \, \hat{e}_1 \tc e_3 &= k_1 \, \hat{e}_4 \tc e_1+k_2 \, \hat{e}_1 \tc e_3 \,,\el
 K \, \hat{e}_1 \tc e_4 &= k_1 \, \hat{e}_6 \tc e_1+k_2 \, \hat{e}_1 \tc e_4 \,,\el
 K \, \hat{e}_2 \tc e_1 &= (1-q^2 k_3) \, \hat{e}_2 \tc e_1+k_5 \, \hat{e}_8 \tc e_1+\big(\tfrac{1}{q}+q\big) k_3 \, \hat{e}_1 \tc e_2 + k_4 \, \hat{e}_6 \tc e_3-\tfrac{k_4}{q} \, \hat{e}_4 \tc e_4 \,,\el
 K \, \hat{e}_2 \tc e_2 &= \big(1+\tfrac{1}{q^2}\big) k_3 \, \hat{e}_3 \tc e_1 + \big(1-\tfrac{k_3}{q^2}\big) \, \hat{e}_2 \tc e_2-\tfrac{k_5}{q^2} \, \hat{e}_8 \tc e_2 - \tfrac{k_4}{q^2} \, \hat{e}_7 \tc e_3 + \tfrac{k_4}{q^3} \, \hat{e}_5 \tc e_4 \,,\el
 K \, \hat{e}_2 \tc e_3 &= \tfrac{k_1}{q} \, \hat{e}_5 \tc e_1 + k_1 \, \hat{e}_4 \tc e_2+k_2 \, \hat{e}_2 \tc e_3 \,,\el
 K \, \hat{e}_2 \tc e_4 &= \tfrac{k_1}{q} \, \hat{e}_7 \tc e_1 + k_1 \, \hat{e}_6 \tc e_2+k_2 \, \hat{e}_2 \tc e_4 \,,\el
 K \, \hat{e}_3 \tc e_1 &= (1-(1+q^2) k_3) \, \hat{e}_3 \tc e_1+k_3 \, \hat{e}_2 \tc e_2+k_5 \, \hat{e}_8 \tc e_2+k_4 \, \hat{e}_7 \tc e_3-\tfrac{k_4}{q} \, \hat{e}_5 \tc e_4 \,,\el
 K \, \hat{e}_3 \tc e_2 &= \, \hat{e}_3 \tc e_2 \,,\el
 K \, \hat{e}_3 \tc e_3 &= k_1 \, \hat{e}_5 \tc e_2+k_2 \, \hat{e}_3 \tc e_3 \,,\el
 K \, \hat{e}_3 \tc e_4 &= k_1 \, \hat{e}_7 \tc e_2+k_2 \, \hat{e}_3 \tc e_4 \,,\el
 K \, \hat{e}_4 \tc e_1 &= k_6 \, \hat{e}_4 \tc e_1+\big(\tfrac{1}{q}+q\big) k_{11} \, \hat{e}_1 \tc e_3 \,,\el
 K \, \hat{e}_4 \tc e_2 &= k_{12} \, \hat{e}_5 \tc e_1+\big(k_6-\tfrac{k_{12}}{q}\big) \, \hat{e}_4 \tc e_2+q k_{11} \, \hat{e}_2 \tc e_3-\tfrac{k_{13}}{q} \, \hat{e}_8 \tc e_3 \,,\el
 K \, \hat{e}_4 \tc e_3 &= k_7 \, \hat{e}_4 \tc e_3 \,,\el
 K \, \hat{e}_4 \tc e_4 &= k_8 \, \hat{e}_2 \tc e_1+k_{10} \, \hat{e}_8 \tc e_1 - \tfrac{1+q^2}{q^3} k_8 \, \hat{e}_1 \tc e_2 + k_9 \, \hat{e}_6 \tc e_3+\big(k_7-\tfrac{k_9}{q}\big) \, \hat{e}_4 \tc e_4 \,,\el
 K \, \hat{e}_5 \tc e_1 &= (k_6-q k_{12}) \, \hat{e}_5 \tc e_1+k_{12} \, \hat{e}_4 \tc e_2+k_{11} \, \hat{e}_2 \tc e_3+k_{13} \, \hat{e}_8 \tc e_3 \,,\el
 K \, \hat{e}_5 \tc e_2 &= k_6 \, \hat{e}_5 \tc e_2+\big(\tfrac{1}{q}+q\big) k_{11} \, \hat{e}_3 \tc e_3 \,,\el
 K \, \hat{e}_5 \tc e_3 &= k_7 \, \hat{e}_5 \tc e_3 \,,\el
 K \, \hat{e}_5 \tc e_4 &= \big(1+\tfrac{1}{q^2}\big) k_8 \, \hat{e}_3 \tc e_1-\tfrac{k_8}{q^2} \, \hat{e}_2 \tc e_2 + k_{10} \, \hat{e}_8 \tc e_2+k_9 \, \hat{e}_7 \tc e_3+\big(k_7-\tfrac{k_9}{q}\big) \, \hat{e}_5 \tc e_4 \,,\el
 K \, \hat{e}_6 \tc e_1 &= k_6 \, \hat{e}_6 \tc e_1+\big(\tfrac{1}{q}+q\big) k_{11} \, \hat{e}_1 \tc e_4 \,,\el
 K \, \hat{e}_6 \tc e_2 &= k_{12} \, \hat{e}_7 \tc e_1+\big(k_6-\tfrac{k_{12}}{q}\big) \, \hat{e}_6 \tc e_2+q k_{11} \, \hat{e}_2 \tc e_4-\tfrac{k_{13}}{q} \, \hat{e}_8 \tc e_4 \,,\el
 K \, \hat{e}_6 \tc e_3 &= -q k_8 \, \hat{e}_2 \tc e_1-q k_{10} \, \hat{e}_8 \tc e_1+\big(1+\tfrac{1}{q^2}\big) k_8 \, \hat{e}_1 \tc e_2+(k_7-q k_9) \, \hat{e}_6 \tc e_3+k_9 \, \hat{e}_4 \tc e_4 \,,\el
 K \, \hat{e}_6 \tc e_4 &= k_7 \, \hat{e}_6 \tc e_4 \,,\el
 K \, \hat{e}_7 \tc e_1 &= (k_6-q k_{12}) \, \hat{e}_7 \tc e_1+k_{12} \, \hat{e}_6 \tc e_2+k_{11} \, \hat{e}_2 \tc e_4+k_{13} \, \hat{e}_8 \tc e_4 \,,\el
 K \, \hat{e}_7 \tc e_2 &= k_6 \, \hat{e}_7 \tc e_2 + \big(\tfrac{1}{q}+q\big) k_{11} \, \hat{e}_3 \tc e_4 \,,\el
 K \, \hat{e}_7 \tc e_3 &= -(\tfrac{1}{q}+q) k_8 \, \hat{e}_3 \tc e_1+\tfrac{k_8}{q} \, \hat{e}_2 \tc e_2 - q k_{10} \, \hat{e}_8 \tc e_2+(k_7-q k_9) \, \hat{e}_7 \tc e_3+k_9 \, \hat{e}_5 \tc e_4 \,,\el
 K \, \hat{e}_7 \tc e_4 &= k_7 \, \hat{e}_7 \tc e_4 \,,\el
 K \, \hat{e}_8 \tc e_1 &= k_{14} \, \hat{e}_2 \tc e_1+k_{16} \, \hat{e}_8 \tc e_1-\tfrac{1+q^2}{q^3} k_{14} \, \hat{e}_1 \tc e_2 + k_{15} \, \hat{e}_6 \tc e_3-\tfrac{k_{15}}{q} \, \hat{e}_4 \tc e_4 \,,\el
 K \, \hat{e}_8 \tc e_2 &= \big(1+\tfrac{1}{q^2}\big) k_{14} \, \hat{e}_3 \tc e_1-\tfrac{k_{14}}{q^2} \, \hat{e}_2 \tc e_2 + k_{16} \, \hat{e}_8 \tc e_2+k_{15} \, \hat{e}_7 \tc e_3-\tfrac{k_{15}}{q} \, \hat{e}_5 \tc e_4 \,,\el
 K \, \hat{e}_8 \tc e_3 &= -q k_{17} \, \hat{e}_5 \tc e_1+k_{17} \, \hat{e}_4 \tc e_2+k_{18} \, \hat{e}_8 \tc e_3 \,,\el
 K \, \hat{e}_8 \tc e_4 &= -q k_{17} \, \hat{e}_7 \tc e_1+k_{17} \, \hat{e}_6 \tc e_2+k_{18} \, \hat{e}_8 \tc e_4 \,.
\end{align}
The reflection coefficients of $K_q^{Ba}$ are
\begin{align}
k_1 &=  \sqrt{\tfrac{q}{1+q^2}}\frac{q(U^4-1)(\xi +x^{-})}{x_B-x^{-}}\frac{V}{U}\frac{\gamma_B}{\gamma}\,,\el
k_2 &=  \frac{q^2 (x_B+\xi )-U^2(\xi +x^{+})}{q^2 U^2 \left(x_B-x^{-}\right)}\,,\el
k_3 &=  \frac{U^4-1}{1+q^2}\frac{\xi +x^{-}}{x_B-x^{-}}\left[q^2+\frac{1}{U{}^2z}\frac{x_B+\xi -U^2(\xi +x^{+})}{(x_B+\xi ) x^{+}}\frac{1+\xi  x^{+}}{\chi+x^{-}}\right]\,,\el
k_4 &=  \frac{k_{13}}{1+\chi x^{+}} \left[\left[\frac{1}{x_B}+\xi -\frac{\xi }{U^2 z}\right] x^{+}-\frac{1}{U^2 z}\right] \frac{\,\gamma\,}{\ul{\gamma}}\,,\el
k_5 &=  \alpha \frac{U^4-1}{1+q^{-2}}\frac{ x^{+}-\kappa(x^{+})- z^{-1}(q^2-1)(\xi + x^{-})}{\chi+x^{-}}\frac{\xi +x^{-}}{x_B-x^{-}}\frac{V^2-U^{-2}}{\gamma^2}\,,\el
k_6 &=  \left[z\,U^2\frac{x_B+\xi }{x^{-}-x_B}+ \frac{\xi +x^{-}}{x^{-}-x_B}\right]\frac{\,\ul{\gamma}\,}{\gamma}\,,\el
k_7 &= \frac{U^2 (\xi +x^{-})+(x_B+\xi ) z}{x^{-}-x_B}\frac{\,\ul{\gamma}\,}{\gamma} \,,\el
k_8 &= \sqrt{\tfrac{q}{1+q^2}} \frac{V}{zU}\frac{x_B+\xi -U^2(\xi +x^{+})}{\kappa(x^{-})-x_B}\frac{U^2(\xi +x^{-})+z(x_B+\xi)}{x^{-}-x_B}\frac{1+\xi x^{+}}{x_B+\xi }\frac{\kappa(x^{-})-x^{-}}{x^{+} \left(\chi+x^{-}\right)}\frac{\ul{\gamma}\gamma_B}{q \alpha }\,,\el
k_9 &=  q z\frac{x^{+}}{x^{-}}\frac{x^{-}-\kappa(x^{-})}{x_B-x^{-}}\left[\frac{\left(\frac{1}{x_B}+\xi \right) x^{+}-\frac{1+\xi  x^{+}}{U^2 z}}{q^2 U{}^2(1+\chi x^{+})}\frac{U^2 (\xi +x^{-})+z(x_B+\xi)}{\chi+ x^{-}}-1\right]\frac{\,\ul{\gamma}\,}{\gamma}\,,\el
k_{10} &=  -\frac{k_1 (\xi +x^{+}) \left(q^2 U^2 \left(1+\xi  x^{+}\right)-(x_B+\xi ) V^2 x^{-}\right) \ul{\gamma}}{q^4 \left(1+x_B \xi +(x_B+\xi ) x^{-}\right) x^{+} \gamma}\,,\el
k_{11} &=  \sqrt{\tfrac{q}{1+q^2}}\frac{U}{q \xi  V}\frac{x_B+\xi }{x^{-}-x_B}\left[\frac{z}{\kappa(x^{-})}+\frac{V^2}{x^{+}}\right] \frac{\ul{\gamma}}{\gamma_B}\,,\el
k_{12} &= \sqrt{\tfrac{q}{1+q^2}} \frac{1}{U V}\left[ \frac{q k_{11} (x^{-}-x^{+}) \gamma_B}{ x^{-} \gamma}+\frac{k_{13} \gamma \gamma_B}{ q \alpha }\right] \,,\el
k_{13} &=  \sqrt{\tfrac{q}{1+q^2}}\frac{ \alpha  (x^{-}-x^{+}) \left(x^{+}-\kappa(x^{+})\right)}{q\text{  }U V\left(\chi+ x^{-}\right) x^{-}}\frac{U^2 (\xi +x^{-})+(x_B+\xi ) z}{x^{-}-x_B}\frac{\ul{\gamma}}{\gamma{}^2\gamma_B}\,,\el
k_{14} &=  \frac{q^2 \left(x^{+}+\xi \right)-\left(\kappa(x^{+})+\xi \right)}{(1+q^2)(x_B-\kappa(x^{-}))} \frac{U^2 (\xi +x^{-})+(x_B+\xi ) z}{V^2(\chi+x^{-})(x^{-}-x_B)}\left[\frac{1}{x^{-}}-\frac{1}{\kappa(x^{-})}\right]\frac{\ul{\gamma}{}^2}{\alpha }\,,\el
k_{15} &= - \sqrt{\tfrac{q}{1+q^2}}\frac{U}{q \xi  V}\frac{x_B+\xi }{x^{-}-x_B}\left[\frac{z}{\kappa(x^{-})}+\frac{V^2}{x^{+}}\right]\left[\frac{\xi +x^{-}}{\chi+x^{-}}-\frac{U^2 z}{x_B}\frac{1+x_B \xi}{\chi+x^{-}}\right]\frac{\ul{\gamma}{}^2}{\gamma\gamma_B}\,,\el
k_{16} &= \alpha  k_{14} V^2 \frac{(\xi +x^{+}) (x^{+}-x^{-})}{\left(1+\xi  x^{+}\right) \gamma^2}+ k_{15} \sqrt{\tfrac{1+q^2}{q^3}}\frac{V}{U} \frac{\xi +x^{+}}{x_B+\xi}\frac{ \gamma_B}{\gamma}+k_6\frac{\xi +x^{+}}{\xi +\kappa(x^{+})} \frac{\,\ul{\gamma}\,}{\gamma}\,,\el
k_{17} &=  \sqrt{\tfrac{q}{1+q^2}}\frac{q^{-2} (\xi +x^{+})+(x_B+\xi ) z}{U V x^{-} (x^{-}-x_B)}\frac{1+\xi  x^{+}}{x_B+\xi }\frac{\kappa(x^{-})-x^{-}}{\chi+x^{-}}\frac{\gamma_B\ul{\gamma}{}^2}{q \alpha  \gamma}\,,\el
k_{18} &=  \frac{q^{-2} (\xi +x^{+})+(x_B+\xi ) z}{x^{-}-x_B}\left[\frac{\xi +x^{-}}{\chi+x^{-}}-\frac{ U^2 z}{q^{2}x_B}\frac{1+x_B \xi}{\chi+x^{-}}\right]\frac{\ul{\gamma}^2}{\gamma^2}\,,
\end{align}
here $\,\chi = \frac{1+\xi x_B }{x_B+\xi}\,$.


\paragraph{Reflection matrix $K^{Ab}_q$}

\begin{align}
 K \, e_1 \tc \hat{e}_1 &= \, e_1 \tc \hat{e}_1 \,,\el
 K \, e_1 \tc \hat{e}_2 &= \big(1+\tfrac{1}{q^2}\big) k_5 \, e_2 \tc \hat{e}_1 + \big(1-\tfrac{k_5}{q^2}\big) \, e_1 \tc \hat{e}_2 - \tfrac{k_7}{q^2} \, e_4 \tc \hat{e}_4 + \tfrac{k_7}{q^3} \, e_3 \tc \hat{e}_6 - \tfrac{k_6}{q^2}\, e_1 \tc \hat{e}_8 \,,\el
 K \, e_1 \tc \hat{e}_3 &= q k_5 \, e_2 \tc \hat{e}_2 + \big(1-k_5+\tfrac{k_5}{q^2}\big) \, e_1 \tc \hat{e}_3 - \tfrac{k_7}{q} \, e_4 \tc \hat{e}_5 + \tfrac{k_7}{q^2} \, e_3 \tc \hat{e}_7 - \tfrac{k_6}{q} \, e_2 \tc \hat{e}_8 \,,\el
 K \, e_1 \tc \hat{e}_4 &= \big(\tfrac{1}{q}+q\big) k_9 \, e_3 \tc \hat{e}_1 + k_1 \, e_1 \tc \hat{e}_4 \,,\el
 K \, e_1 \tc \hat{e}_5 &= q k_9 \, e_3 \tc \hat{e}_2+k_8 \, e_2 \tc \hat{e}_4 + \big(k_1-\tfrac{k_8}{q}\big) \, e_1 \tc \hat{e}_5-\tfrac{k_{10}}{q} \, e_3 \tc \hat{e}_8 \,,\el
 K \, e_1 \tc \hat{e}_6 &= \big(\tfrac{1}{q}+q\big) k_9 \, e_4 \tc \hat{e}_1 + k_1 \, e_1 \tc \hat{e}_6 \,,\el
 K \, e_1 \tc \hat{e}_7 &= q k_9 \, e_4 \tc \hat{e}_2+k_8 \, e_2 \tc \hat{e}_6+\big(k_1-\tfrac{k_8}{q}\big) \, e_1 \tc \hat{e}_7-\tfrac{k_{10}}{q} \, e_4 \tc \hat{e}_8 \,,\el
 K \, e_1 \tc \hat{e}_8 &= -(1+q^2) k_2 \, e_2 \tc \hat{e}_1+k_2 \, e_1 \tc \hat{e}_2+k_4 \, e_4 \tc \hat{e}_4-\tfrac{k_4}{q} \, e_3 \tc \hat{e}_6 + k_3 \, e_1 \tc \hat{e}_8 \,,\el
 K \, e_2 \tc \hat{e}_1 &= (1-(1+q^2) k_5) \, e_2 \tc \hat{e}_1+k_5 \, e_1 \tc \hat{e}_2+k_7 \, e_4 \tc \hat{e}_4-\tfrac{k_7}{q} \, e_3 \tc \hat{e}_6 + k_6 \, e_1 \tc \hat{e}_8 \,,\el
 K \, e_2 \tc \hat{e}_2 &= (1-q^2 k_5) \, e_2 \tc \hat{e}_2+\big(\tfrac{1}{q}+q\big) k_5 \, e_1 \tc \hat{e}_3 + k_7 \, e_4 \tc \hat{e}_5-\tfrac{k_7}{q} \, e_3 \tc \hat{e}_7 + k_6 \, e_2 \tc \hat{e}_8 \,,\el
 K \, e_2 \tc \hat{e}_3 &= \, e_2 \tc \hat{e}_3 \,,\el
 K \, e_2 \tc \hat{e}_4 &= k_9 \, e_3 \tc \hat{e}_2+(k_1-q k_8) \, e_2 \tc \hat{e}_4+k_8 \, e_1 \tc \hat{e}_5+k_{10} \, e_3 \tc \hat{e}_8 \,,\el
 K \, e_2 \tc \hat{e}_5 &= \big(\tfrac{1}{q}+q\big) k_9 \, e_3 \tc \hat{e}_3 + k_1 \, e_2 \tc \hat{e}_5 \,,\el
 K \, e_2 \tc \hat{e}_6 &= k_9 \, e_4 \tc \hat{e}_2+(k_1-q k_8) \, e_2 \tc \hat{e}_6+k_8 \, e_1 \tc \hat{e}_7+k_{10} \, e_4 \tc \hat{e}_8 \,,\el
 K \, e_2 \tc \hat{e}_7 &= \big(\tfrac{1}{q}+q\big)k_9 \, e_4 \tc \hat{e}_3 + k_1 \, e_2 \tc \hat{e}_7 \,,\el
 K \, e_2 \tc \hat{e}_8 &= -q^2 k_2 \, e_2 \tc \hat{e}_2+\big(\tfrac{1}{q}+q\big) k_2 \, e_1 \tc \hat{e}_3 + k_4 \, e_4 \tc \hat{e}_5-\tfrac{k_4}{q} \, e_3 \tc \hat{e}_7 + k_3 \, e_2 \tc \hat{e}_8 \,,\el
 K \, e_3 \tc \hat{e}_1 &= k_{12} \, e_3 \tc \hat{e}_1+k_{11} \, e_1 \tc \hat{e}_4 \,,\el
 K \, e_3 \tc \hat{e}_2 &= k_{12} \, e_3 \tc \hat{e}_2+\tfrac{k_{11}}{q} \, e_2 \tc \hat{e}_4 + k_{11} \, e_1 \tc \hat{e}_5 \,,\el
 K \, e_3 \tc \hat{e}_3 &= k_{12} \, e_3 \tc \hat{e}_3+k_{11} \, e_2 \tc \hat{e}_5 \,,\el
 K \, e_3 \tc \hat{e}_4 &= k_{13} \, e_3 \tc \hat{e}_4 \,,\el
 K \, e_3 \tc \hat{e}_5 &= k_{13} \, e_3 \tc \hat{e}_5 \,,\el
 K \, e_3 \tc \hat{e}_6 &= -(1+q^2) k_{14} \, e_2 \tc \hat{e}_1+k_{14} \, e_1 \tc \hat{e}_2+k_{16} \, e_4 \tc \hat{e}_4+\big(k_{13}-\tfrac{k_{16}}{q}\big) \, e_3 \tc \hat{e}_6+k_{15} \, e_1 \tc \hat{e}_8 \,,\el
 K \, e_3 \tc \hat{e}_7 &= -q^2 k_{14} \, e_2 \tc \hat{e}_2 + \big(\tfrac{1}{q}+q\big) k_{14} \, e_1 \tc \hat{e}_3 + k_{16} \, e_4 \tc \hat{e}_5 + \big(k_{13}-\tfrac{k_{16}}{q}\big) \, e_3 \tc \hat{e}_7+k_{15} \, e_2 \tc \hat{e}_8 \,,\el
 K \, e_3 \tc \hat{e}_8 &= -q k_{17} \, e_2 \tc \hat{e}_4+k_{17} \, e_1 \tc \hat{e}_5+k_{18} \, e_3 \tc \hat{e}_8 \,,\el
 K \, e_4 \tc \hat{e}_1 &= k_{12} \, e_4 \tc \hat{e}_1+k_{11} \, e_1 \tc \hat{e}_6 \,,\el
 K \, e_4 \tc \hat{e}_2 &= k_{12} \, e_4 \tc \hat{e}_2+\tfrac{k_{11}}{q} \, e_2 \tc \hat{e}_6 + k_{11} \, e_1 \tc \hat{e}_7 \,,\el
 K \, e_4 \tc \hat{e}_3 &= k_{12} \, e_4 \tc \hat{e}_3+k_{11} \, e_2 \tc \hat{e}_7 \,,\el
 K \, e_4 \tc \hat{e}_4 &= (q+q^3) k_{14} \, e_2 \tc \hat{e}_1-q k_{14} \, e_1 \tc \hat{e}_2+(k_{13}-q k_{16}) \, e_4 \tc \hat{e}_4+k_{16} \, e_3 \tc \hat{e}_6-q k_{15} \, e_1 \tc \hat{e}_8 \,,\el
 K \, e_4 \tc \hat{e}_5 &= q^3 k_{14} \, e_2 \tc \hat{e}_2-(1+q^2) k_{14} \, e_1 \tc \hat{e}_3+(k_{13}-q k_{16}) \, e_4 \tc \hat{e}_5+k_{16} \, e_3 \tc \hat{e}_7-q k_{15} \, e_2 \tc \hat{e}_8 \,,\el
 K \, e_4 \tc \hat{e}_6 &= k_{13} \, e_4 \tc \hat{e}_6 \,,\el
 K \, e_4 \tc \hat{e}_7 &= k_{13} \, e_4 \tc \hat{e}_7 \,,\el
 K \, e_4 \tc \hat{e}_8 &= -q k_{17} \, e_2 \tc \hat{e}_6+k_{17} \, e_1 \tc \hat{e}_7+k_{18} \, e_4 \tc \hat{e}_8 \,.
\end{align}
The reflection coefficients of $K_q^{Ab}$ are
\begin{align}
k_1 &= \frac{(x_B+\xi )-U^4 (\xi +x^{-})}{U^2(x_B-x^{-})} \,,\el
k_2 &= \frac{q^{-1}k_4+qU^{-2}k_9}{\alpha \sqrt{1+q^2}}\frac{U V \gamma \gamma_B}{U^2V^2-1} \,,\el
k_3 &= \frac{k_{10} -\alpha U^2 k_4 }{\alpha q\sqrt{1+q^2} \,(1-U^2V^2)}\frac{V}{U} - k_{18}\frac{\gamma}{zU^2} \,,\el
k_4 &=  \frac{\alpha q^2 k_{17}}{\gamma \ul{\gamma}}\frac{\kappa(x^{+})x^{-}}{(U^2-V^2){}^{-1}} - k_{16} \frac{q^2 UV}{z\sqrt{1+q^2}}\frac{x^{-}+\xi }{x_B+\xi}\frac{\gamma_B}{\ul{\gamma}} \,,\el
k_5 &=  \frac{U^2-U{}^{-2}}{1+q^2}\bigg[\frac{V^2U^2}{\xi q^2}\frac{1+x_B \xi }{x_B-x^{-}}\frac{q^2-1}{1-x_B x^{+}}\left[\frac{ 1+\xi x^{+}}{U^2}-\frac{x^{+} (x_B+\xi )}{U^2-V^2}\right] \el
 & \hspace{30mm} + q \frac{\xi +x^{+}}{x_B-x^{-}} \left[1-\frac{V^2-U^2}{q \kappa(x^{-})x^{+}}+\frac{q V^2(\xi ^2-1)}{(x_B+\xi)(\xi +x^{+})}\right]\bigg] ,\el
k_7 &= -q \alpha \sqrt{1+q^2} k_5 \frac{1-U^2V^2}{U V\gamma \gamma_B} + \frac{q^3 \alpha}{\sqrt{1+q^2}} \frac{1-U^2V^2}{U V \gamma \gamma_B} - \frac{q^3 \alpha k_{12}}{\sqrt{1+q^2}}\frac{1-U^{-2}V^2}{U V \ul{\gamma} \gamma_B}\,,\el
k_6 &=  \tfrac{1}{\sqrt{1+q^2}}\frac{U V}{U^2-V^2}\frac{x_B+\xi}{\xi +x^{+}}\frac{\gamma}{\gamma_B}(k_{10} U^2-k_7) \,,\el
k_8 &=  \frac{q}{q^2+1}\frac{(1+x_B^2+2 x_B \xi)x^{+}}{(x_B-x^{-})(1-x_B x^{+})} \frac{U^4-1}{V^2-U^2}\frac{V^2}{U^2} \,,\el
k_9 &=  \frac{U^2-U{}^{-2}}{\sqrt{1+q^2}}\frac{V}{U}\frac{\xi +x^{+}}{x_B-x^{-}} \,,\el
k_{10} &= \alpha \frac{U^2-U^{-2}}{\sqrt{q^2+1}}\frac{V}{U}\frac{\xi +x^{+}}{x_B-x^{-}}\left[1+\frac{(1-q^4) x_B (1+\xi x^{+})}{\xi (1-x_B x^{+})}\right] ,\el
k_{11} &= \sqrt{\tfrac{1}{1+q^2}}\frac{U^4-1}{U^2-V^2}\frac{V}{U}\frac{x_B+\xi }{x_B-x^{-}} \,,\el
k_{12} &=  -\left[\frac{\xi +x^{-}}{x_B-x^{-}}+zU^2\frac{x_B+\xi}{x_B-x^{-}}\right] ,\el
k_{13} &=  -\frac{\xi+x^{+}}{q(x_B-x^{-})}-z\frac{x_B+\xi }{x_B-x^{-}} \,,\el
k_{14} &=  \frac{U^2-U^{-2}}{\alpha q^4\sqrt{1+q^2}}\frac{V}{U}\frac{1+x_B \xi}{x_B-x^{-}}\left[\frac{U^2V^2}{U^2-V^2}\left[\frac{(q^2-1)(1+\xi x^{+})}{\xi(1-x_B x^{+})}-\frac{1}{x_B}\right]-\frac{z}{q^2 x_B}\frac{x_B+\xi}{\xi +x^{-}}\right] ,\el
k_{15} &=  \frac{q^{-1}}{U^2V^2-1}\left[k_{16} \frac{U V}{\sqrt{1+q^2}} + \alpha^{-1}k_{10}\right] ,\el
k_{16} &=  \frac{U^4-1}{x_B-x^{-}} \frac{V^2 }{U^2}\left[z(x_B+\xi)\frac{q^2(x^{+}+\xi)-(1+x_B \xi ) x^{+}}{q^3 \xi (1-x_B x^{+})}-(x^{+}+\xi)\right] ,\el
k_{17} &=  \frac{1}{\alpha q\sqrt{1+q^2}}\frac{V}{U}\frac{q z(x_B+\xi )+ (\xi +x^{+})}{\kappa(x^{+})(1-x_B x^{+})}\frac{x^{+}-\kappa(x^{+})}{x_B-x^{-}} \,,\el
k_{18} &=  \frac{(x_B+\xi)x^{-}+(\xi  +x^{+})\frac{1}{\kappa(x^{+})}}{x_B-x^{-}}\frac{q z(x_B+\xi )+(\xi +x^{+})}{1-x_B x^{+}}\frac{V^2}{q} \,.
\end{align}


\global\long\def\nlin#1#2{\href{http://xxx.lanl.gov/abs/nlin/#2}{\tt nlin.#1/#2}}
\global\long\def\hepth#1{\href{http://xxx.lanl.gov/abs/hep-th/#1}{\tt hep-th/#1}}
\global\long\def\condmat#1{\href{http://xxx.lanl.gov/abs/cond-mat/#1}{\tt cond-mat/#1}}
\global\long\def\arXivid#1{\href{http://arxiv.org/abs/#1}{\tt arXiv:#1}}
\global\long\def\xmath#1{\href{http://arxiv.org/abs/math/#1}{\tt math/#1}}


\begin{thebibliography}{plain}

\enlargethispage{0.1in}

\bibitem{review}
  N.~Beisert et~al.,
  \textit{Review of AdS/CFT Integrability: An Overview}, 
  Lett.~Math.~Phys.~99:3--32, 2012, 
  [\arXivid{1012.3982}].

\bibitem{BeisertFundamental}
  N.~Beisert, 
  \textit{The SU$(2|2)$ Dynamic $S$-matrix},
  Adv.\ Theor.\ Math.\ Phys.\  {12} (2008) 945, 
  [\arXivid{0511082}].

\bibitem{BAnalytic}
  N.~Beisert, 
  \textit{The Analytic Bethe Ansatz for a Chain with Centrally Extended su$(2|2)$ symmetry}, 
  J.\ Stat.\ Mech.\ {0701} (2007) P01017, 
  [\nlin{0610017}].

\bibitem{AFPZ}
  G.~Arutyunov, S.~Frolov, J.~Plefka and M.~Zamaklar, 
  \textit{The off-shell Symmetry Algebra of the Light-cone AdS(5) x S**5 Superstring},
  J.\ Phys.\ A {40} (2007) 3583,
  [\arXivid{hep-th/0609157}].

\bibitem{BeisertYangian}
  N.~Beisert, 
  \textit{The $S$-matrix of AdS / CFT and Yangian symmetry}, 
  PoS {SOLVAY} (2006) 002, 
  [\arXivid{0704.0400}].

\bibitem{AFS}
  G.~Arutyunov, S.~Frolov and M.~Staudacher, 
  \textit{Bethe Ansatz For Quantum Strings}, 
  JHEP {0410} (2004) 016, 
  [\arXivid{0406256}] $\bullet$
%
  N.~Beisert and M.~Staudacher, 
  \textit{Long-range $psu(2,2|4)$ Bethe Ansatze for gauge theory and strings}, Nucl.\ Phys.\ B {727} (2005) 1 
  [\arXivid{0504190}].

\bibitem{MdLrmat}
  M.~de Leeuw, 
  \textit{bound-states, Yangian Symmetry and Classical r-matrix for the $\ads$ Superstring}, 
  JHEP {0806} (2008) 085,
  [\arXivid{0804.1047}].

\bibitem{AFBound}
  G.~Arutyunov and S.~Frolov, 
  \textit{The $S$-matrix of String Bound States}, 
  Nucl.~Phys.~B 804:90-143 (2008), 
  [\arXivid{0803.4323}].

\bibitem{deLeeuw:2008ye}
  M.~de Leeuw, 
  \textit{The Bethe Ansatz for AdS(5) x S**5 bound-states}, 
  JHEP {0901} (2009) 005,
  [\arXivid{0809.0783}].

\bibitem{ALT}
  G.~Arutyunov, M.~de Leeuw and A.~Torrielli, 
  \textit{The bound-state $S$-matrix for $\ads$ Superstring}, 
  Nucl.~Phys.~B 819:319-350, 2009, 
  [\arXivid{0902.0183}].


\bibitem{Spectral}
  N.~Gromov, V.~Kazakov and P.~Vieira, 
  \textit{Exact Spectrum of Anomalous Dimensions of Planar N=4 Supersymmetric Yang-Mills Theory},
  Phys.\ Rev.\ Lett.\  {103} (2009) 131601, 
  [\arXivid{0901.3753}] $\bullet$
%
  D.~Bombardelli, D.~Fioravanti and R.~Tateo, 
  \textit{Thermodynamic Bethe Ansatz for planar AdS/CFT: A Proposal},
  J.\ Phys.\ A {42} (2009) 375401, 
  [\arXivid{0902.3930}] $\bullet$
%
  G.~Arutyunov and S.~Frolov, 
  \textit{Thermodynamic Bethe Ansatz for the AdS(5) x S(5) Mirror Model},
  JHEP {0905} (2009) 068, 
  [\arXivid{0903.0141}].
  

\bibitem{Wilson}
  N.~Drukker, \textit{Integrable Wilson loops},
  [\arXivid{1203.1617}] $\bullet$
%
  D.~Correa, J.~Maldacena and A.~Sever, 
  \textit{The quark anti-quark potential and the cusp anomalous dimension from a TBA equation},
  [\arXivid{1203.1913}].
  

\bibitem{McGreevy:2000cw}
  J.~McGreevy, L.~Susskind and N.~Toumbas,
  \textit{Invasion of the giant gravitons from Anti-de Sitter space},
  JHEP {0006} (2000) 008,
  [\hepth{0003075}].

\bibitem{OpenB}
  %
  A.~Hashimoto, S.~Hirano and N.~Itzhaki, 
  {\it Large branes in AdS and their field theory dual}, 
  JHEP 08 (2000) 051,
  [\hepth{0008016}] $\bullet$
  %
  V.~Balasubramanian and A.~Naqvi, 
  {\it Giant gravitons and a correspondence principle}, 
  Phys.~Lett.~B 528, 111 (2002),
  [\hepth{0111163}] $\bullet$ 
  %
  V.~Balasubramanian et~al.,
  {\it Giant Gravitons in Conformal Field Theory},
  JHEP 0204 (2002) 034,
  [\hepth{0107119}] $\bullet$
  %
  A.~Dabholkar and S.~Parvizi, 
  {\it Dp-branes in pp-wave background}
  [\hepth{0203231}] $\bullet$
  %
  D.~Berenstein et~al., 
  {\it Open strings on plane waves and their Yang-Mills duals}, 
  [\hepth{0203249}] $\bullet$ 
  %
  P.~Lee and J.~W.~Park,
  {\it Open strings in PP-wave background from defect conformal field theory}, 
  Phys.~Rev.~D~67 (2003) 026002,
  [\hepth{0203257}] $\bullet$
  %
  B.~Stefanski, jr., 
  {\it Open spinning strings}, 
  JHEP 0403, 057 (2004),
  [\hepth{0312091}] $\bullet$
  %
  B.~Stefanski, jr., 
  {\it Open string plane-wave light-cone superstring field theory},
  Nuc.~Phys.~B 666, 71 (2003),
  [\hepth{0304114}] $\bullet$
  %
  B.~Chen, X.~J.~Wang and Y.~S.~Wu, 
  {\it Integrable open spin chain in super Yang-Mills and the plane-wave / SYM duality},
  JHEP 0402, 029 (2004),
  [\hepth{0401016}] $\bullet$
  %
  B.~Chen, X.~J.~Wang and Y.~S.~Wu, 
  {\it Open spin chain and open spinning string}, 
  Phys.~Lett.~B 591, 170 (2004),
  [\hepth{0403004}].
  
  
\bibitem{DOz}
  A.~Dekel and Y.~Oz, 
  \textit{Integrability of Green-Schwarz Sigma Models with Boundaries}, 
  JHEP 1108 (2011) 004, 
  [\arXivid{1106.3446}].
  

\bibitem{HM}
  D.~M.~Hofman and J.~Maldacena, 
  \textit{Reflecting magnons},
  JHEP 0711 (2007) 050, 
  [\arXivid{0708.2272}].

\bibitem{D3refs}
  D.~Berenstein and S.~E.~Vazquez,
  {\it Integrable open spin chains from giant gravitons},
  JHEP 0506, 059 (2005),
  [\hepth{0501078}] $\bullet$  
%
  D.~Berenstein, D.~H.~Correa and S.~E.~Vazquez,
  {\it A Study of open strings ending on giant gravitons, spin chains and integrability},
  JHEP {\bf 0609} (2006) 065,
  [\hepth{0604123}] $\bullet$  
%
  L.~Palla, 
  \textit{Issues on magnon reflection}, 
  Nucl.~Phys.~B~808 (2009) 205-223, 
  [\arXivid{0807.3646}] $\bullet$
%
  C.~Ahn, D.~Bak and S.~J.~Rey, 
  \textit{Reflecting Magnon bound-states},
  JHEP 0804 (2008) 050, 
  [\arXivid{0712.4144}] $\bullet$
%
  Z.~Bajnok and L.~Palla, 
  \textit{Boundary finite size corrections for multiparticle states and planar AdS/CFT},
  JHEP 01 (2011) 011, 
  [\arXivid{1010.5617}] $\bullet$
%
  R.~I.~Nepomechie,
  \textit{Revisiting the Y=0 open spin chain at one loop}, JHEP {1111} (2011) 069,
  [\arXivid{1109.4366}] $\bullet$
%
  Z.~Bajnok, R.~I.~Nepomechie, L.~Palla and R.~Suzuki,
  \textit{Y-system for Y=0 brane in planar AdS/CFT},
  [\arXivid{1205.2060}].
  
 
\bibitem{D7D5refs}
  A.~Karch and L.~Randall,
  \textit{Open and closed string interpretation of SUSY CFT's on branes with boundaries},
  JHEP {0106} (2001) 063,
  [\hepth{0105132}] $\bullet$ 
%
  O.~DeWolfe, D.~Z.~Freedman and H.~Ooguri,
  \textit{Holography and defect conformal field theories},
  Phys.\ Rev.\ D {66} (2002) 025009, 
  [\hepth{0111135}] $\bullet$  
%
  A.~Karch and E.~Katz,
  \textit{Adding flavor to AdS / CFT},
  JHEP {0206} (2002) 043,
  [\hepth{0205236}] $\bullet$
%
  O.~DeWolfe and N.~Mann, 
  {\it Integrable open spin chains in defect conformal field theory}, 
  JHEP 0404, 035 (2004),
  [\hepth{0401041}] $\bullet$ 
%
  Y.~Susaki, Y.~Takayama and K.~Yoshida, 
  {\it Open semiclassical strings and long defect operators in AdS/dCFT correspondence},
  Phys.~Rev.~D~71 (2005) 126006,
  [\hepth{0410139}] $\bullet$  
%
  D.~H.~Correa, V.~Regelskis and C.~A.~S.~Young, 
  \textit{Integrable achiral D5-brane reflections and asymptotic Bethe equations}, 
  J.\ Phys.\ A {44} (2011) 325403, 
  [\arXivid{1105.3707}].

\bibitem{Kruczenski:2003be}
  M.~Kruczenski, D.~Mateos, R.~C.~Myers and D.~J.~Winters,
  \textit{Meson spectroscopy in AdS/CFT with flavor},
  JHEP {0307} (2003) 049,
  [\hepth{0304032}].

\bibitem{CY}
  D.~H.~Correa and C.~A.~S.~Young, \textit{Reflecting magnons from D7 and D5 branes}, 
  J.~Phys.~A 41 (2008) 455401, [\arXivid{0808.0452}].

\bibitem{MR1}
  N.~MacKay and V.~Regelskis, 
  \textit{On the reflection of magnon bound-states}, 
  JHEP 1008 (2010) 055, 
  [\arXivid{1006.4102}].

\bibitem{LeeuwThesis}
  M.~de Leeuw, 
  \textit{The $S$-matrix of the $AdS_5 x S^5$ superstring}, 
  [\arXivid{1007.4931}] $\bullet$
%
  A.~Torrielli, \textit{Yangians, $S$-matrices and AdS/CFT},
  J.\ Phys.\ A {44} (2011) 263001,
  [\arXivid{1104.2474}].
  
\bibitem{Sk}
  E.~K.~Sklyanin, 
  {\it Boundary conditions for integrable quantum systems}, 
  J.~Phys.~{A 21} (1988) 2375.

\bibitem{TwYangians}  
  G.~I.~Olshanskii, 
  {\it Twisted Yangians and infinite dimensional classical Lie algebras}, 
  Lecture Notes in Mathematics {1510}, Proceedings Leningrad (1990), Springer $\bullet$ 
%
  A.~I.~Molev and E.~Ragoucy,
  \textit{Representations of reflection algebras,}
  Rev.~Math.~Phys.~{14} (2002) 317, 
  [\xmath{0107213}] $\bullet$
%
  A.~Mudrov,
  \textit{Reflection equation and twisted Yangians,}
  J.~Math.~Phys.~{48} (2007) 093501, 
  [\xmath{0612737}] $\bullet$
%
  N.~Guay, X.~Ma,
  {\it Twisted Yangians, twisted quantum loop algebras and affine Hecke algebras of type BC},
  2011.  

\bibitem{DMS}
  G.~W.~Delius, N.~J.~MacKay and B.~J.~Short,
  \textit{Boundary remnant of Yangian symmetry and the structure of rational reflection matrices}, 
  Phys.\ Lett.\  B {522} (2001) 335,
  [Erratum-ibid.\  B {524} (2002) 401],
  [\arXivid{hep-th/0109115}].

\bibitem{MacKay:2002at}
  N.~J.~MacKay,
  {\it Rational K matrices and representations of twisted Yangians},
  J.\ Phys.\ A {35} (2002) 7865,
  [\arXivid{math/0205155}].

\bibitem{MK}
  N.~J.~MacKay, 
  \textit{Introduction to Yangian symmetry in integrable field theory}, 
  Int.~J.~Mod.~Phys.~A 20 (2005) 71897218, 
  [\hepth{0409183}].

\bibitem{VR}
  V.~Regelskis, 
  \textit{Reflection algebras for SL(2) and GL(1$|$1)}, 
  [\arXivid{1206.6498}].

\bibitem{YangianY0}
  C.~Ahn and R.~I.~Nepomechie, 
  \textit{Yangian symmetry and bound-states in AdS/CFT boundary scattering}, 
  JHEP 1005 (2010) 016, 
  [\arXivid{1003.3361}] $\bullet$
%
  N.~MacKay and V.~Regelskis, 
  \textit{Yangian Symmetry of the Y=0 Maximal Giant Graviton}, 
  JHEP 1012 (2010) 076, 
  [\arXivid{1010.3761}].

\bibitem{Palla1}
  L.~Palla, 
  \textit{Yangian symmetry of boundary scattering in AdS/CFT and the explicit form of bound-state reflection matrices},
  JHEP 0804 (2008) 022, 
  [\arXivid{1102.0122}].

\bibitem{MR3}
  N.~MacKay and V.~Regelskis, 
  \textit{Reflection Algebra, Yangian Symmetry and Bound-States in AdS/CFT}, 
  JHEP {1201} (2012) 134, 
  [\arXivid{1101.6062}].

\bibitem{MR4}
  N.~MacKay and V.~Regelskis,
  \textit{Achiral boundaries and the twisted Yangian of the D5-brane},
  JHEP {1108} (2011) 019, [\arXivid{1105.4128}].

\bibitem{BGM}
  N.~Beisert, W.~Galleas and T.~Matsumoto 
  \textit{A Quantum Affine Algebra for the Deformed Hubbard Chain}, 
  [\arXivid{1102.5700}].

\bibitem{LMR}
  M.~de Leeuw, T.~Matsumoto and V.~Regelskis, 
  \textit{The bound-state $S$-matrix of the Deformed Hubbard Chain}, 
  JHEP {1204} (2012) 021, 
  [\arXivid{1109.1410}].

\bibitem{BK}
  N.~Beisert and P.~Koroteev, 
  \textit{Quantum Deformations of the One-Dimensional Hubbard Model},
  J.~Phys.~A~41 (2008) 255204,
  [\arXivid{0802.0777}].

\bibitem{Hoare:2011fj}
  B.~Hoare and A.~A.~Tseytlin,
  {\it Towards the quantum S-matrix of the Pohlmeyer reduced version of $AdS_5 \times S^5$ superstring theory},
  Nucl.\ Phys.\ B {851} (2011) 161,
  [\arXivid{1104.2423}],
  
\bibitem{HoareBound}
  B.~Hoare, T.~J.~Hollowood and J.~L.~Miramontes,
  {\it Bound States of the q-Deformed $AdS_5 \times S^5$ Superstring S-matrix},
    [\arXivid{1206.0010}].

\bibitem{MN}
  R.~Murgan and R.~Nepomechie, 
  \textit{q-deformed $su(2|2)$ boundary $S$-matrices via the ZF algebra}, 
  JHEP 0806 (2008) 096, 
  [\arXivid{0805.3142}].

\bibitem{LMR2}
  M.~de Leeuw, T.~Matsumoto and V.~Regelskis, 
  \textit{Coideal Quantum Affine Algebra and Boundary Scattering of the Deformed Hubbard Chain}, 
  J.~Phys.~A.~45 (2012) 065205, 
  [\arXivid{1110.4596}].

\bibitem{Le}
  G.~Letzter, 
  \textit{Coideal subalgebras and quantum symmetric pairs,} 
  in ‘New directions in Hopf algebras’, 
  Math.~Sci.~Res.~Inst.~Publ.~{43}, Cambridge Univ.~Press,
  Cambridge (2002) 117, [\xmath{0103228}] $\bullet$
%
  G.~Letzter, 
  \textit{Quantum symmetric pairs and their zonal spherical functions,}
  Transformation Groups {8} (2003) 261, 
  [\xmath{0204103}].
  
\bibitem{MRSHK}
  A.~I.~Molev, E.~Ragoucy and P.~Sorba, 
  \textit{Coideal subalgebras in quantum affine algebras}, 
  Rev.~Math.~Phys.~15 (2003) 789822,
  [\xmath{0208140}] $\bullet$
%
  I.~Heckenberger and S.~Kolb,
  {\it Homogeneous right coideal subalgebras of quantized enveloping algebras},
  [\arXivid{1109.3986}].

\bibitem{Gomez:2006va}
  C.~Gomez and R.~Hernandez,
  \textit{The Magnon kinematics of the AdS/CFT correspondence},
  JHEP {0611} (2006) 021,
  [\arXivid{hep-th/0608029}] $\bullet$
%
  J.~Plefka, F.~Spill and A.~Torrielli,
  \textit{On the Hopf algebra structure of the AdS/CFT $S$-matrix},
  Phys.\ Rev.\ D {74} (2006) 066008,
  [\arXivid{hep-th/0608038}].
  
\bibitem{BMN}
  D.~E.~Berenstein, J.~M.~Maldacena and H.~S.~Nastase,
  \textit{Strings in flat space and pp waves from N=4 Super Yang-Mills},
  AIP Conf.\ Proc.\  {646} (2003) 3,
  [\arXivid{hep-th/0202021}]. 

\bibitem{GZ}
  S.~Ghoshal and A.~B.~Zamolodchikov, 
  {\it Boundary S-matrix and boundary state in two-dimensional integrable quantum field theory}, 
  Int.~J.~Mod.~Phys.~{A 9} (1994) {3841}, 
  [\hepth{9306002}].
    
\bibitem{D2}
  V.~G.~Drinfeld, 
  {\it A new realization of Yangians and quantum affine algebras}, 
  Soviet Math.~Dokl.~36 (1988) 212.

\bibitem{ST}
  F.~Spill and A.~Torrielli, 
  \textit{On Drinfeld's second realization of the AdS/CFT su(2$\vert$2) Yangian}, 
  J.~Geom.~Phys.~ 59 (2009) 489502, 
  [\arXivid{0803.3194}].

\bibitem{ALT2}
  G.~Arutyunov, M.~de~Leeuw and A.~Torrielli, 
  \textit{On Yangian and Long Representations of the Centrally Extended $su(2|2)$ Superalgebra},
  JHEP 1006 (2010) 033, 
  [\arXivid{0912.0209}].

\bibitem{RegelskisSecret}
  V.~Regelskis, 
  \textit{The Secret symmetries of the AdS/CFT reflection matrices}, 
  JHEP {1108} (2011) 006, 
  [\arXivid{1105.4497}].

\bibitem{MMT}
  T.~Matsumoto, S.~Moriyama and A.~Torrielli, 
  \textit{A Secret Symmetry of the AdS/CFT $S$-matrix}, 
  JHEP {0709} (2007) 099, 
  [\arXivid{0708.1285}] $\bullet$
%
  M.~de Leeuw et~al.,
  {\it Secret Symmetries in AdS/CFT}, to appear in Physica Scripta,
  [\arXivid{1204.2366}].

\bibitem{deLeeuwSecret}
  M.~de Leeuw, V.~Regelskis and A.~Torrielli, 
  \textit{The Quantum Affine Origin of the AdS/CFT Secret Symmetry},
  J.\ Phys.\ A {45} (2012) 175202,   
  [\arXivid{1112.4989}].  

\end{thebibliography}
\end{document}